\PassOptionsToPackage{unicode}{hyperref}
\PassOptionsToPackage{hyphens}{url}
\PassOptionsToPackage{dvipsnames,svgnames*,x11names*}{xcolor}
\documentclass[
  12pt,
  letterpaper,
]{article}
\usepackage{lmodern}
\usepackage{setspace}
\usepackage{amssymb,amsmath}
\usepackage{ifxetex,ifluatex}
\ifnum 0\ifxetex 1\fi\ifluatex 1\fi=0 
  \usepackage[T1]{fontenc}
  \usepackage[utf8]{inputenc}
  \usepackage{textcomp} 
\else 
  \usepackage{unicode-math}
  \defaultfontfeatures{Scale=MatchLowercase}
  \defaultfontfeatures[\rmfamily]{Ligatures=TeX,Scale=1}
\fi
\IfFileExists{upquote.sty}{\usepackage{upquote}}{}
\IfFileExists{microtype.sty}{
  \usepackage[]{microtype}
  \UseMicrotypeSet[protrusion]{basicmath} 
}{}
\makeatletter
\@ifundefined{KOMAClassName}{
  \IfFileExists{parskip.sty}{%
    \usepackage{parskip}
  }{
    \setlength{\parindent}{0pt}
    \setlength{\parskip}{6pt plus 2pt minus 1pt}}
}{
  \KOMAoptions{parskip=half}}
\makeatother
\usepackage{xcolor}
\IfFileExists{xurl.sty}{\usepackage{xurl}}{} 
\IfFileExists{bookmark.sty}{\usepackage{bookmark}}{\usepackage{hyperref}}
\hypersetup{
  pdftitle={Dead Reckoning},
  colorlinks=true,
  linkcolor=black,
  filecolor=Maroon,
  citecolor=black,
  urlcolor=blue,
  pdfcreator={LaTeX via pandoc}}
\usepackage[margin=1in]{geometry}
\usepackage{longtable,booktabs}
\usepackage{etoolbox}
\makeatletter
\patchcmd\longtable{\par}{\if@noskipsec\mbox{}\fi\par}{}{}
\makeatother
\IfFileExists{footnotehyper.sty}{\usepackage{footnotehyper}}{\usepackage{footnote}}
\makesavenoteenv{longtable}
\usepackage{graphicx}
\makeatletter
\def\maxwidth{\ifdim\Gin@nat@width>\linewidth\linewidth\else\Gin@nat@width\fi}
\def\maxheight{\ifdim\Gin@nat@height>\textheight\textheight\else\Gin@nat@height\fi}
\makeatother
\setkeys{Gin}{width=\maxwidth,height=\maxheight,keepaspectratio}
\makeatletter
\def\fps@figure{htbp}
\makeatother
\usepackage{mathptmx}   
\usepackage{graphicx}
\usepackage{booktabs}
\usepackage{caption}
\usepackage{placeins}

\title{Dead Reckoning}
\usepackage{etoolbox}
\makeatletter
\providecommand{\subtitle}[1]{
  \apptocmd{\@title}{\par {\large #1 \par}}{}{}
}
\makeatother
\subtitle{Counting Your Customers Who Never Say Goodbye}
\author{Karl T. Ulrich\\
The Wharton School, University of Pennsylvania\\
ulrich@wharton.upenn.edu}
\date{Working draft v4, July 6, 2026}

\begin{document}
\maketitle

\setstretch{1.5}
\vspace{-1.2em}
\setstretch{1.0}

\textbf{Abstract.} Firms in non-contractual commerce face the challenge of knowing how many customers they actually have because their customers can stop buying without ever saying they have left. Buy-Till-You-Die models address this problem by estimating each customer's probability of being alive, a quantity usually called P(alive) and used in essentially every major software tool for generating dashboards, defining churn, calculating customer equity, and valuing the enterprise. We show that this practice confounds two distinct quantities. Within the beta-geometric family of models, P(alive) is the infinite-horizon limit of an observable family of finite-horizon probabilities of repeat purchase. Every finite-horizon estimate, such as the probability of repeat purchase within 12 months, is a forecast of a verifiable event. The limit at infinite time can only be estimated by extrapolation, a customer count obtained by dead reckoning. The implied customer count is therefore only partially identified: realized returners are the lower bound, and estimation conventions determine the reported point estimate above that. The consequences are significant. In our empirical setting, on a seven-year transaction panel for 31,683 unique customers, model specifications with nearly identical observable forecasts estimate the number of ``alive'' customers at values ranging from 3,654 to 27,734, a difference of 7.6x. A default weighting parameter set in customer base software swings the count 42 percent. Five years of subsequent purchases further falsify the maximum-likelihood count from below. The main patterns we find can be replicated on the CDNOW benchmark, with a 2.4x spread in point estimates of the size of the customer base. Most of what practice calls miscalibration is instead a category error: summed P(alive) overshoots the realized eighteen-month returners by 2.25x, while the same model's own eighteen-month forecast errs by just 1.18x. The remedy is to report and maintain an auditable horizon count, say the number of actual repeat customers within 18 months, estimate return probabilities at a stated horizon, audit them across scoring dates and horizons, recalibrate them as customer cohorts drift, and report the total customer count, if it must be reported at all, as a confidence interval rather than a point.

\small

\textbf{Acknowledgments and disclosures.} The author thanks Peter S. Fader for generous conversations and for pointers to antecedent work. The author is a co-founder of MakerStock, the company that provided the empirical data for this study. This research received no external funding.

\normalsize
\setstretch{1.5}

\newpage

\hypertarget{introduction}{%
\subsection{1. Introduction}\label{introduction}}

Dead reckoning is how ships navigated between star sightings: they extrapolated from the last known position using assumed speed and heading. Its defining feature is that error is invisible from inside the ship and accumulates until the next external fix. A similar logic underlies one of the most widely reported quantities in customer-base analysis: the expected number of ``live'' customers. This paper argues that existing customer-base models can generate substantial errors in that estimate, characterizes those errors, and proposes practical remedies.

The setting is non-contractual commerce, where customers typically do not announce their departure (Ascarza, Netzer, and Hardie 2018). A silent customer may have defected or may simply be between purchases, yet many managerial questions depend on resolving that ambiguity. For nearly four decades, the canonical solution has been the Buy-Till-You-Die (BTYD) framework, beginning with the Pareto/NBD model of Schmittlein, Morrison, and Colombo (1987) and its estimation-friendly successor, the BG/NBD model of Fader, Hardie, and Lee (2005). These models translate each customer's silence into a posterior probability of being ``alive,'' a quantity now generated by major customer analytics platforms and used in retention thresholds, churn-risk dashboards, customer-equity calculations, and customer-based corporate valuation of noncontractual firms (McCarthy and Fader 2018). This paper asks whether any transaction record and model can answer the basic question of how many customers such a firm actually has, and what firms should report when it cannot.

The key identity is simple. In the BG/NBD family, ``alive'' means ``will purchase again eventually,'' because dropout occurs only at a purchase. P(alive) is therefore the finite-horizon question ``will this customer order again within H months?'' with H taken to infinity. Every finite-H member of this family is a forecast of a verifiable event, gradeable by anyone willing to wait H months. But no one can wait forever. The infinite-horizon customer count is reached only by extrapolation, and the extrapolation depends on parametric assumptions about unobserved heterogeneity that the likelihood only weakly constrains. The sum of P(alive) over the customer base is therefore the one member of an observable family that its own data can never fully grade. It is a dead-reckoned position.

This distinction matters because many decisions use probability levels, not merely rankings or aggregate transaction forecasts. Acquisition-versus-retention budgets depend on how many recoverable customers the organization has; ``40,000'' and ``18,000'' prescribe different strategies. Many marketing metrics place an active-customer count in the denominator. Customer equity and customer-based valuations weight per-customer values by aliveness. At the individual level, breakeven winback rules spend c when return probability times margin exceeds c, and profit-based churn targeting formalizes exactly such rules (Lemmens and Gupta 2020). A score can rank customers well while overstating probability levels threefold; such a score may pass targeting audits while approving spend on customers who are gone. Aggregate purchase forecasts can survive miscalibrated levels, but counting customers, valuing customers, and computing unit economics cannot.

\hypertarget{identification-evidence-and-remedy}{%
\subsubsection{1.1 Identification, Evidence, and Remedy}\label{identification-evidence-and-remedy}}

This research is organized around three questions. First, what does the reported customer count mean, and can any finite transaction record accurately deliver it? Second, how large are the practical consequences, both in an operating firm and on the field's canonical dataset? Third, what should firms report instead, and how should that number be maintained as conditions drift? The results are: the reported count is the infinite-horizon limit of an auditable family of forecasts and is only partially identified by transaction data; the consequences are large, with counts varying by a factor of 7.6 across interchangeable specifications on our panel and by 42 percent from a software default alone; and firms should report a calibrated horizon count, maintained dynamically and audited on a grid of scoring dates and horizons.

The identification problem can be stated without model machinery. A transaction record reveals, to a close approximation, the product of two quantities: how many customers remain alive and how fast the living ones purchase. Three hundred living customers buying at rate four and two hundred living customers buying at rate six both generate twelve hundred purchases, closely similar histories, and nearly the same likelihood. Timing information, especially recency, loosens this equivalence but does not eliminate it. In Abe's (2009) simulations from a correctly specified model, purchase rates are recovered at a correlation of 0.80 with truth, while dropout rates are recovered at 0.18.\footnote{Recovery correlations as reported in the simulation appendix of the working-paper version (Abe 2008, CIRJE-F-537); to be confirmed against the typeset article at proof stage.} Jerath, Fader, and Hardie (2011) likewise showed that models with nearly indistinguishable fit imply dramatically different death-related quantities, concluding that fit ``is not sufficient to judge the suitability of a model for inferential purposes.'' The likelihood surface has a ridge, and whatever breaks the tie along it---model structure, a prior, or, as we show, the default numerical penalty in standard fitting software---selects the reported customer count on grounds the data cannot audit.

Three empirical results give this argument practical force. First, the count behaves like a dial. On a seven-year, two-regime transaction panel from an operating company using BTYD models, six specifications with statistically indistinguishable discrimination and near-identical eighteen-month forecasts place the same 31,683 customers' alive count anywhere from 3,654 to 27,734. Within one specification, the software-default ridge penalty moves the count from 3,660 to 5,211, a 42 percent increase, while changing observable forecasts by less than rounding; in a simulation study against known truth, the same default inflates the count by 53 percent. A replication on the CDNOW data shows the same pattern: counts from 8,446 to 19,981 on identical AUC, with the software default setting worth 22 percent.

Second, patience adjudicates only the lower bound. At a vintage with sixty months of subsequent observation, 419 realized returners falsify the maximum-likelihood estimate of 210 within thirty months, while the trajectory closest to reality belonged to a model parameterized by an arbitrary penalty factor. The count is only partially identified: bounded below by accumulating observed returners and bounded above only weakly, with every estimation convention selecting an unauditable point inside the interval.

Third, much of the practical error is a category error rather than a model failure. Scored against eighteen-month outcomes, summed P(alive) overshoots by 2.25x; for BG/NBD it overshoots by 11.99x and produces worse-than-random ranking. The same fitted model's own eighteen-month forecast is off by only 1.18x. No re-estimation separates these numbers. They are two outputs of one model, differing only by the horizon.

The constructive program is a navigation discipline: take fixes, correct course, and never mistake extrapolation for observation. The statistical components are standard tools applied to a newly articulated challenge rather than new methods. We grade the model's native horizon forecasts with reliability diagrams and proper scoring rules (Brier 1950; Dimitriadis, Gneiting, and Jordan 2021) on a grid of scoring vintages by horizons, out of time by construction. The grid matters because drift can make the field's conventional single calibration/holdout split misleading: a misspecified tail whose bias cancels a regime change can appear well calibrated in one cell while being falsified across the grid.

For course correction, we develop a dynamic calibration layer that exploits an empirical regularity in our panel: acquisition cohorts mature along a stable shape while differing in level, the same structure actuaries exploit in loss-development triangles. The layer estimates the shape from pooled history and the current level from the freshest partially observed cohorts. The underlying logic, estimating a long-horizon quantity from short-horizon signals, has a causal-inference cousin in the surrogate-index targeting of Huang and Ascarza (2024); here it is used for calibration maintenance rather than treatment assignment. In a thirteen-quarter out-of-time backtest, the layer repairs a failure mode we document for static maps, which can be worse than no correction when trained on a distorted vintage. It is competitive overall, performs best in the drift-regime quarters where correction matters, outperforms cheaper structural alternatives in those quarters, and produces a drift alarm that flagged a regime change two quarters before full-horizon validation could.

The resulting reporting rule is deliberately conservative. The calibrated horizon count should be the operating KPI for the organization. The infinite-horizon customer count, if reported at all, should be reported as an identified interval.

\hypertarget{scope-related-work-and-organization}{%
\subsubsection{1.2 Scope, Related Work, and Organization}\label{scope-related-work-and-organization}}

The scope of our claims is bounded. The limit identity is exact for the BG family and becomes an inequality for the Pareto/NBD, whose P(alive) bounds every observable finite-horizon member from above. The category error is known in principle; the contribution here is to measure its magnitude and document it in the wild. Our survey of software documentation and practitioner materials finds aliveness probabilities exposed by every major package, routinely thresholded and dashboarded as churn risk, and glossed in official documentation as the probability that the customer will purchase again. Explicit summation into active-customer counts appears mainly in valuation-adjacent framing rather than in tutorials. Finally, this paper does not challenge the aggregate transaction forecasting that constitutes most BTYD use. The results concern decisions based on probability levels.

The BTYD tradition models a noncontractual customer as a latent two-process system: purchase while alive, die silently, never return. The Pareto/NBD and BG/NBD anchor the family; the modified BG/NBD adds a dropout opportunity at time zero so one-time buyers can be dead (Batislam, Denizel, and Filiztekin 2007; equivalently Hoppe and Wagner 2007); the periodic death opportunity model decouples death from purchases (Jerath, Fader, and Hardie 2011); the Pareto/GGG generalizes purchase timing (Platzer and Reutterer 2016; see also Reutterer, Platzer, and Schröder 2021); and hierarchical Bayes and covariate extensions follow (Abe 2009; Gopalakrishnan, Bradlow, and Fader 2017; Dew and Ansari 2018; Dew, Ansari, and Li 2020; Bachmann, Meierer, and Näf 2021). The validation norms were set by Fader, Hardie, and Lee (2005): calibration/holdout splits scored by aggregate tracking plots, conditional expectations of transactions by past frequency, and individual-level correlations of predicted with realized purchase counts. These are demanding tests of the transaction process, but every element grades transactions. P(alive), introduced as an inferential by-product, is exposed and consumed but rarely scored as a probability. The field is aware of the underlying looseness, as Jerath et al.'s warning and Abe's recovery gap show. The most direct antecedent of our reframing is Fader, Hardie, and Shang (2010), who state that P(alive) ``is a prediction of something that is, by definition, unobservable,'' making it ``impossible to directly assess its validity''; who insist that ``alive'' (an unobservable state) not be conflated with ``active'' (observable behavior); and who propose the conditional penetration, the discrete-time analogue of our \(R_H\), as the observable companion measure. Wünderlich (2015) develops the same instinct in a hierarchical Bayes setting, working with the probability of zero purchases over a finite window and modeling individual-level seasonality. What the literature has lacked is what follows from taking these warnings to their conclusion: the identification result, the calibration audit, and the maintenance layer this paper supplies, together with the documentation that practice adopted the P(alive) output while shelving the accompanying advice.

Three adjacent streams come close without meeting the target. Wübben and von Wangenheim (2008) validated Pareto/NBD activity classification against holdout purchase incidence and found that a recency heuristic matched the model; their hit-rate design dichotomizes the probability at a cutoff and therefore tests classification, not probability levels. The machine-learning CLV stream predicts fixed-horizon activity directly and evaluates by discrimination (Martínez et al.~2020; Neslin et al.~2006), with Chamberlain et al.~(2017) additionally recalibrating churn probabilities at a noncontractual retailer and Padilla and Ascarza (2021) bringing deep generative machinery to thin first-purchase histories. But redefining the target dissolves the latent aliveness construct rather than auditing it. Finally, the probability-calibration toolkit is mature (Cox 1958; Zadrozny and Elkan 2002; Gneiting and Raftery 2007; Guo et al.~2017; Dimitriadis, Gneiting, and Jordan 2021), as are its uses in goodness-of-fit testing (Hosmer and Lemeshow 1980; Spiegelhalter 1986; Andrews 1988), clinical validation and recalibration of latent-risk models against fixed-horizon events (Van Calster et al.~2016, 2019; Davis et al.~2017), censoring-aware survival evaluation (Graf et al.~1999; Demler, Paynter, and Cook 2015), and actuarial reserving through stable-shape cohort logic (Bornhuetter and Ferguson 1972). We import these tools as they are. What has been missing is the target: to our knowledge, no prior work scores the probability calibration of BTYD-implied aliveness quantities, interprets the results as evidence about identification of the latent structure, or maintains the correction dynamically.

The remainder of the paper is organized as follows. Section 2 develops the framework: the three quantities, the limit identity, and partial identification of the customer count. Section 3 presents the method: the calibration audit and recalibration layers, including a head-to-head against structural alternatives. Section 4 reports the empirical evidence: the field study, the CDNOW replication, and the simulations. Section 5 develops the managerial implications, including the randomized design for the firm's forthcoming reactivation campaign. Section 6 concludes. An appendix collects formal statements.

\hypertarget{framework-aliveness-horizon-returns-and-identification}{%
\subsection{2. Framework: Aliveness, Horizon Returns, and Identification}\label{framework-aliveness-horizon-returns-and-identification}}

\hypertarget{three-quantities}{%
\subsubsection{2.1 Three quantities}\label{three-quantities}}

For a customer with sufficient statistics (\(x\), \(t_x\), \(T\)) at vintage v (the scoring date), a fitted BTYD model delivers three outputs that practice tends to treat interchangeably: the aliveness probability \(A\) = P(alive \(\mid\) \(x\), \(t_x\), \(T\)); the expected number of transactions in the window, \(E[X_H]\), where \(X_H\) counts purchases in \((v, v+H]\); and the horizon-return probability \(R_H\) = P(\(X_H \geq 1\)), the continuous-time counterpart of the conditional penetration that Fader, Hardie, and Shang (2010) recommend reporting alongside P(alive) precisely because it is observable. All three are closed-form functions of the same statistics. For the BG family, conditional on being alive at \(v\) the purchase rate has a gamma posterior, and since dropout occurs only at purchases, zero purchases means zero dropout opportunities, giving

\[R_H = A \times \left[1 - \left(\frac{\alpha + T}{\alpha + T + H}\right)^{r+x}\right].\]

\(R_H\) is \(A\) times a purchase-timing factor, so \(A \geq R_H\) always, with the wedge largest for slow buyers. \(E[X_H]\) is a count not a probability. Summed over customers, \(R_H\) yields \(N_H\), the expected number of customers who return within the window; \(E[X_H]\) yields an order count; and \(A\) yields the number dashboards typically report as the customer count, which we write \(N_\infty\) for reasons clarified in the next subsection. The field's validation battery grades models by order counts, not number of customers, a metric on which they score well. Dashboards and thresholds typically treat \(A\) as if it were \(R_H\).

\begin{longtable}[]{@{}ll@{}}
\toprule
\begin{minipage}[b]{0.47\columnwidth}\raggedright
Symbol\strut
\end{minipage} & \begin{minipage}[b]{0.47\columnwidth}\raggedright
Meaning\strut
\end{minipage}\tabularnewline
\midrule
\endhead
\begin{minipage}[t]{0.47\columnwidth}\raggedright
\(x\), \(t_x\), \(T\)\strut
\end{minipage} & \begin{minipage}[t]{0.47\columnwidth}\raggedright
repeat frequency, recency, observation length (literature convention)\strut
\end{minipage}\tabularnewline
\begin{minipage}[t]{0.47\columnwidth}\raggedright
\(v\), \(H\)\strut
\end{minipage} & \begin{minipage}[t]{0.47\columnwidth}\raggedright
vintage (scoring date); horizon in months; the audit grid is indexed \((v, H)\)\strut
\end{minipage}\tabularnewline
\begin{minipage}[t]{0.47\columnwidth}\raggedright
\(A\)\strut
\end{minipage} & \begin{minipage}[t]{0.47\columnwidth}\raggedright
aliveness probability, P(alive)\strut
\end{minipage}\tabularnewline
\begin{minipage}[t]{0.47\columnwidth}\raggedright
\(R_H\)\strut
\end{minipage} & \begin{minipage}[t]{0.47\columnwidth}\raggedright
probability of at least one order in \((v, v+H]\)\strut
\end{minipage}\tabularnewline
\begin{minipage}[t]{0.47\columnwidth}\raggedright
\(E[X_H]\)\strut
\end{minipage} & \begin{minipage}[t]{0.47\columnwidth}\raggedright
expected orders in the window\strut
\end{minipage}\tabularnewline
\begin{minipage}[t]{0.47\columnwidth}\raggedright
\(N_H\), \(N_\infty\)\strut
\end{minipage} & \begin{minipage}[t]{0.47\columnwidth}\raggedright
expected returners within \(H\) (sum of \(R_H\)); the dead-reckoned customer count (sum of \(A\))\strut
\end{minipage}\tabularnewline
\begin{minipage}[t]{0.47\columnwidth}\raggedright
\(B\)\strut
\end{minipage} & \begin{minipage}[t]{0.47\columnwidth}\raggedright
aggregate bias, \(N_H\) over realized returners\strut
\end{minipage}\tabularnewline
\begin{minipage}[t]{0.47\columnwidth}\raggedright
\(g\), \(D\)\strut
\end{minipage} & \begin{minipage}[t]{0.47\columnwidth}\raggedright
calibration map; drift statistic (level tilt), \(D = 1\) when the regime matches training\strut
\end{minipage}\tabularnewline
\begin{minipage}[t]{0.47\columnwidth}\raggedright
\(\rho\)\strut
\end{minipage} & \begin{minipage}[t]{0.47\columnwidth}\raggedright
ridge penalty constant in estimation software\strut
\end{minipage}\tabularnewline
\bottomrule
\end{longtable}

That substitution overstates activity by construction, before any question of misspecification arises, and Section 4 measures the resulting wedge in an operating system.

That the substitution is common practice rather than one firm's idiosyncrasy is documentable. All four major implementations (the Python lifetimes package, the R BTYD and CLVTools packages, and PyMC-Marketing) expose per-customer P(alive) as a headline output; practitioner materials routinely threshold it (0.5, 0.8, and 0.9 rules appear in published tutorials), report its complement as a churn-risk score in production retention systems, and consume it as a horizon quantity inside CLV shortcut formulas; and the category error appears inside official documentation, where PyMC-Marketing's quickstart titles a P(alive) trajectory plot ``probability customer will purchase again.'' Explicit summation into an active-customer count is the rarer documented pattern, appearing mainly in valuation-adjacent framing (Rust, Lemon, and Zeithaml's 2004 description of the Pareto/NBD as a model ``for estimating the number of active customers'') and in production systems like the one studied here. Notably, our survey found no source in the software or practitioner literature that warns against reading P(alive) as a horizon-return probability; the caveats that exist concern one-time-buyer artifacts. The academic literature does contain the warning, in Fader, Hardie, and Shang (2010); sixteen years on, none of the tooling that operationalized their model family carries it. An appendix table gives sources.

\hypertarget{the-limit-and-partial-identification}{%
\subsubsection{2.2 The limit, and partial identification}\label{the-limit-and-partial-identification}}

In the BG family a customer alive at \(v\) purchases again with probability one given unlimited time, so letting \(H\) grow in the display above,

\[A = \lim_{H \to \infty} R_H, \qquad \text{and hence} \qquad N_\infty = \lim_{H \to \infty} N_H.\]

Aliveness is the last member of an observable family, the member at H equals forever. (For the Pareto/NBD, where death can occur between purchases, the limit sits strictly below \(A\), so the aliveness probability bounds every observable member from above; the arguments below apply with that substitution.) Each finite-H member is a proper forecast of a verifiable event; the limit's value is carried by the parametric tail of the heterogeneity distributions, which is the ridge direction the likelihood does not price.

The correct formal statement, developed in the appendix, is that \(N_\infty\) is partially identified. The data supply a hard, accumulating lower bound: customers observed to return after \(v\) were alive at \(v\), so realized eventual returners bound the count from below, and the bound tightens as long as anyone keeps returning. The data constrain the count from above only through the flattening of the observed return curve, a weak and slowly arriving restriction. Inside the bounds, structure, penalty, and prior select a point on grounds the data cannot audit; hierarchical Bayes implementations are prior-dominated in exactly this direction for exactly this reason. The identification question here differs from the data-fusion setting of McCarthy and Oblander (2021), who ask which customer-base quantities can be recovered when only aggregated or selectively disclosed data are available; our point is that the count remains set-identified even with the complete individual transaction log in hand. The practical injunction follows: report the interval if a count must be reported, and report a finite-H member of the family, which the data discipline, as the operating number.

\hypertarget{choosing-the-horizon}{%
\subsubsection{2.3 Choosing the horizon}\label{choosing-the-horizon}}

Anchoring on \(R_H\) requires choosing H, trading coverage against verification latency. On our panel an eighteen-month window captures 75 percent of the customers who will have returned by month sixty, in a ratio stable across cohorts, and validating an H-month forecast requires vintages at least H months old, so longer horizons mean staler calibration and slower drift detection. We operate at H = 18 with H = 24 as a reported sensitivity, and we note the discipline the limit identity makes obvious: the calibration target, the validation grid, and the KPI must use the same H. The gap between the 18- and 24-month counts is itself a diagnostic for slow-replenishment business. Seasonality deserves explicit mention. Whole-year horizons integrate the outcome window over the seasonal cycle exactly; the eighteen-month default spans six calendar quarters, which dampens without eliminating seasonal composition, and the residual appears as level movement that the quarterly-refit dynamic layer absorbs. Seasonality on the fitting side remains an omitted covariate for every model in the comparison, which is one more reason to grade deployed forecasts with an audit that is agnostic to what the model omits.

\hypertarget{method-auditing-and-recalibrating-the-horizon-forecast}{%
\subsection{3. Method: Auditing and Recalibrating the Horizon Forecast}\label{method-auditing-and-recalibrating-the-horizon-forecast}}

\hypertarget{the-calibration-audit}{%
\subsubsection{3.1 The calibration audit}\label{the-calibration-audit}}

The audit machinery is standard and we state it briefly, with formal details in the appendix. For each fully observed cell (vintage \(v\), horizon \(H\)), we grade the model's native \(R_H\) forecasts, fit only on data through \(v\), against realized outcomes in \((v, v+H]\), using isotonic (CORP) reliability diagrams with consistency bands (Dimitriadis, Gneiting, and Jordan 2021), aggregate bias \(B\) (predicted over realized returners), the Brier score and its decomposition, and AUC for discrimination. Under correct specification, true parameters, and a stable environment, these forecasts are calibrated at every cell (our simulations verify the machinery passes when nothing is wrong), so a calibration failure rejects the joint hypothesis, in the classical goodness-of-fit sense running from Cox (1958) to Andrews (1988). Three properties of the test discipline its use. It has power only out of time, since estimation approximately enforces calibration on the training window. It is nearly powerless against the tail: observationally equivalent specifications share finite-horizon forecasts, so a calibrated cell certifies neither the factorization nor the count, and only the deep-horizon members of the grid, where realized returners accumulate against each model's asymptote, discipline the tail at all. And within a single cell, structural error and regime change are confounded: the bias at \((v, H)\) sums a structural component that varies with H in a model-specific way and a regime component that varies with calendar time and moves all models together. A single calibration/holdout split, the field's convention, observes one number and cannot decompose it, which is how a misspecified model with compensating bias gets crowned; the grid separates the components by their signatures. Inference respects two dependence structures: outcomes within a vintage share a calendar window, so interval statements are made at the vintage level (moving-block bootstrap over the quarterly series), and cells with incompletely observed windows are either excluded, as here, or require the censoring-aware scores of survival analysis.

\hypertarget{remedies-and-what-each-is-worth}{%
\subsubsection{3.2 Remedies, and what each is worth}\label{remedies-and-what-each-is-worth}}

The audit orders the remedies, and a theme runs through them: every remedy disciplines observable members of the family; none rescues the count, which stays interval-identified regardless.

\textbf{The category fix comes first and costs nothing.} Report the model's own \(R_H\) at a stated horizon instead of \(A\). On our panel this substitution, involving no new estimation, accounts for most of the headline discrepancy (Section 4.2): the same fitted object that overshoots eighteen-month returners by 2.25x when its aliveness output is consumed as an activity count is off by 1.18x on its own eighteen-month forecast. The fix also makes the KPI auditable, which everything else requires.

\textbf{Respecification repairs structural signatures and nothing else.} The plain BG/NBD's forced P(alive) = 1 for one-time buyers produces aliveness scores that are inverted as well as inflated in a one-and-done-heavy base (AUC 0.32: proven repeaters score below single triers), and any descendant repairs this. Beyond that clear case, the equivalence class binds: the hard-core-spike specification and the MBG/NBD fit our data near-identically with coinciding forecasts and, unpenalized, coinciding counts; a profile likelihood over the spike share is nearly flat; and the shared maximum-likelihood tail is falsified by the five-year audit anyway. Nor does decoupling death from purchases resolve it: the periodic death opportunity structure, which removes the buy-die correlation entirely, simply selects yet another point inside the interval (Section 4.3). Getting the structure right by likelihood criteria does not locate the count, and no static specification of any structure stayed calibrated across our vintages.

\textbf{Static recalibration works, with a warning label the literature lacks.} An isotonic map from stated to realized frequencies, fit on a fully observed vintage (Zadrozny and Elkan 2002), preserves ranking, wraps any model, and, applied out of time from the 2023 to the 2024 vintage, lands every model and both scores within 1.06x to 1.13x of realized returners, including the raw BG/NBD aliveness score from 11.99x to 1.29x. But a static map transfers its training regime forward: standing at quarters where the only fully observed vintages were small pandemic-era fits, the freshest-vintage map predicted roughly half of realized returns (bias 0.49 to 0.56) while the raw forecast sat near parity. An unrefreshed calibration layer is not a conservative default; it is a bet that the past persists.

\textbf{The dynamic layer converts drift from nuisance to model, and survives a head-to-head.} Acquisition cohorts on our panel mature along a stable shape while differing in level: normalized to their eighteen-month value, the 2020 through 2024 cohort curves are nearly superimposable (age-to-age development factors \(f_k\), for cohort \(i\) at age \(k\), vary 2 to 6 percent across cohorts through a pandemic and a market expansion), while levels move materially, and a cohort's six-month rate predicts its eighteen-month rate almost exactly (2024 cohort: 18.4 predicted, 18.5 realized). This is the structure of an actuarial loss-development triangle, and the remedy is Bornhuetter and Ferguson's (1972) reserving logic applied to customer return: an isotonic shape pooled over all fully observed vintages, times a level tilt estimated from the freshest partially observed vintages (the triangle's newest diagonal), clipped to the unit interval. In the thirteen-quarter out-of-time backtest the dynamic layer repairs most of the static maps' early failures (0.64 to 0.88 against their 0.49 to 0.64), tracks the raw forecast through the stable mid-sample, and is the best method on the board in the drift-regime quarters (1.11x at the final quarter against 1.18x raw and 1.27x static). Its drift statistic \(D\) (navigators estimating a current's set and drift from recent fixes and applying the correction to subsequent reckoning perform an operation of the same character; the analogy is illustrative, not a derivation) crossed one in mid-2024 and fell to 0.91 by year-end, flagging the regime change two quarters before an eighteen-month validation could see it.

A skeptic should ask whether structural dynamics would do better. To address this question we ran the comparison. Per-acquisition-cohort re-estimation (separate MBG/NBD parameters by cohort, a cheap analogue of cross-cohort changepoint modeling) and rolling-window re-estimation (parameters from the most recent three years of entrants) were scored on the same grid. Across all thirteen quarters the pooled model's raw forecast is the best point performer (mean absolute log bias 0.070, block-bootstrap 95 percent interval 0.024 to 0.122), with rolling-window close behind (0.075) and per-cohort fitting notably worse (0.121, interval 0.084 to 0.164, the price of small-cohort noise); in the four drift-regime quarters the dynamic layer leads everything (0.143 against 0.150 raw, 0.172 rolling, 0.176 per-cohort). The intervals overlap, and thirteen quarters cannot rank the methods definitively; what the comparison establishes is bounded but useful: cheap structural re-parameterization does not dominate a pooled model with a level tilt, it costs more, and it lacks the alarm. The full hierarchical Bayes changepoint machinery was not run and may do better; it would still require this audit to know.

\textbf{The hierarchy.} Report \(R_H\), never \(A\). Respecify when the grid shows a structural signature. Treat any static map as perishable. Prefer the dynamic layer where partial recent outcomes exist. State the count, if it must be stated, as the interval.

\hypertarget{empirical-evidence}{%
\subsection{4. Empirical Evidence}\label{empirical-evidence}}

\hypertarget{setting-and-design}{%
\subsubsection{4.1 Setting and design}\label{setting-and-design}}

Our data come from a U.S. firm, MakerStock, which sells physical goods to institutional customers and consumers. We have every transaction over the company's seven year history up until July 2026, comprising 82,715 qualifying orders from 52,246 customers. Two segments are analyzed separately throughout. The institutional segment (1,272 customers, 5,244 orders) is defined by each geographically distinct site; the consumer segment (50,974 customers at the account grain, 77,406 orders) is roughly 80 percent one-time buyers.

The panel contains four sources of variation a single-firm study rarely offers together: an acute macro shock (the firm's first eighteen months coincide with the COVID-19 pandemic), a chronic composition drift (later cohorts are drawn from harder-to-reach corners of each segment), several distinct go-to-market approaches each sustained for at least six months, and the two-segment contrast under held-fixed operations and calendar. We consider this is a positive feature of the data. One of the challenges in practice is accommodating heterogeneity in customers and shifting regimes.

Models are fit only on data through each scoring vintage and graded on windows the estimation never saw; the grid comprises quarterly vintages (2021-2024) for the backtest, year-end vintages for the model comparison at H = 18 and 24 months, and 2020-2022 vintages with horizons to 60 months for the deep-horizon audit. The firm's own production analytics run an MBG/NBD with a static isotonic layer, and our pipeline reproduces its documented backtest exactly (the 2.14x gap on the 2024 acquisition cohort). The phenomena reported here were discovered in trying to use the values from the MBG/NBD model, and eventually uncovering the category error in the metrics.

\hypertarget{the-category-error}{%
\subsubsection{4.2 The category error}\label{the-category-error}}

At the 2024 year-end vintage (consumer regime, 31,683 customers, 2,319 realized eighteen-month returners), summed P(alive) scored against eighteen-month outcomes overshoots by 2.25x (customer-bootstrap 95 percent interval 2.17 to 2.33) for the MBG/NBD and 11.99x (11.5 to 12.5) for the BG/NBD, whose ranking is also inverted (AUC 0.32). The same fitted MBG/NBD's own eighteen-month forecast is off by 1.18x (1.14 to 1.22) with expected calibration error 0.013 and AUC 0.80. The raw score's reliability curve over-predicts everywhere and is worst in the low deciles (customers scored 0.08 to 0.11 return at 0.015 to 0.025); the native forecast's curve hugs the diagonal with a mild tilt. No re-estimation separates the two panels; they are two outputs of one fitted object. The wedge is definitional in sign (Section 2.1); what these numbers supply is its magnitude in a real organization, and the demonstration that most of what a practitioner would call miscalibration of P(alive) is not miscalibration at all. The institutional segment repeats the pattern at smaller magnitude (raw 1.57x, interval 1.46 to 1.69; native 1.16x, 1.08 to 1.24), consistent with its smaller one-and-done share.

\begin{figure}
\centering
\includegraphics{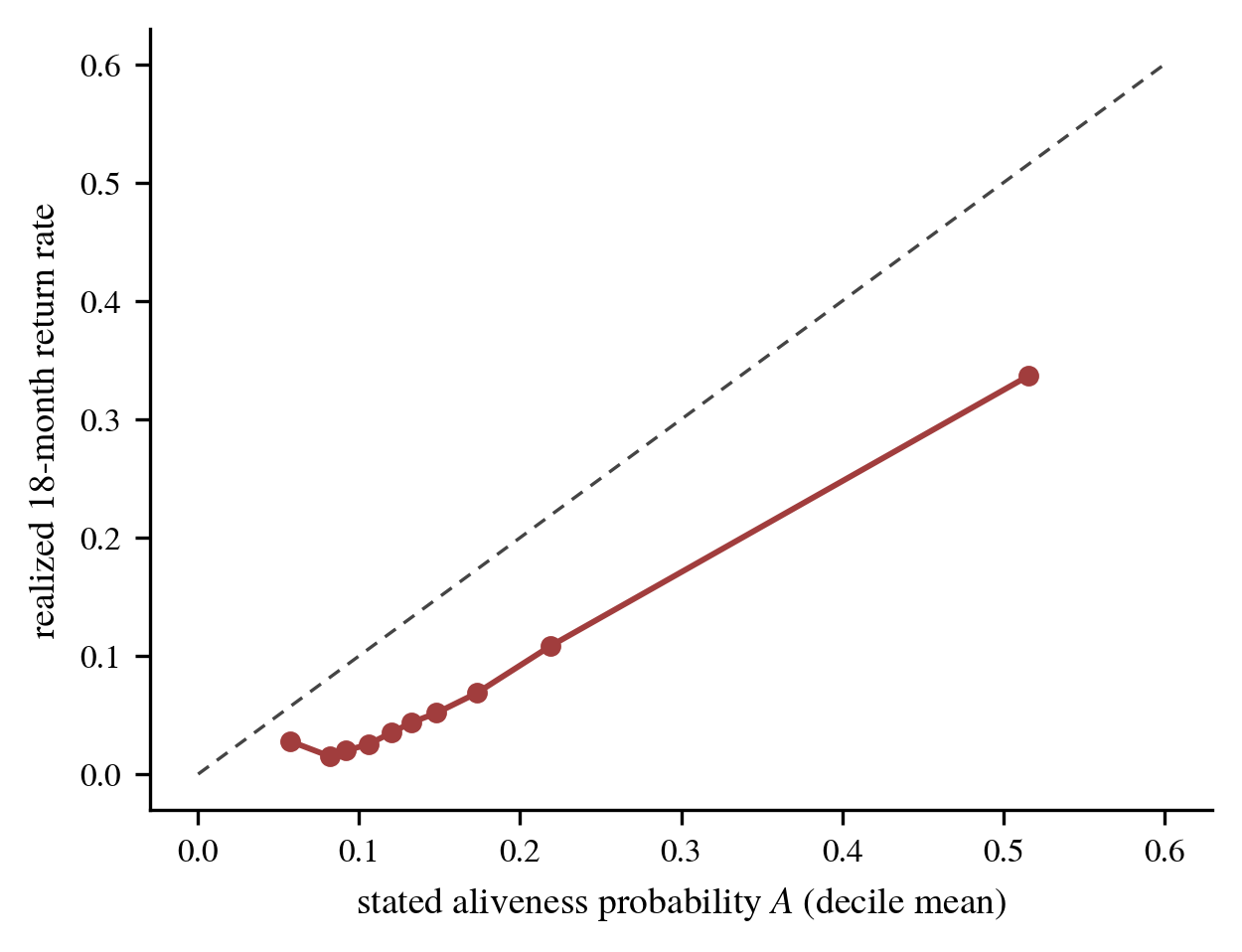}
\caption{The raw aliveness score graded against 18-month outcomes (consumer, 2024 vintage).}
\end{figure}

\begin{figure}
\centering
\includegraphics{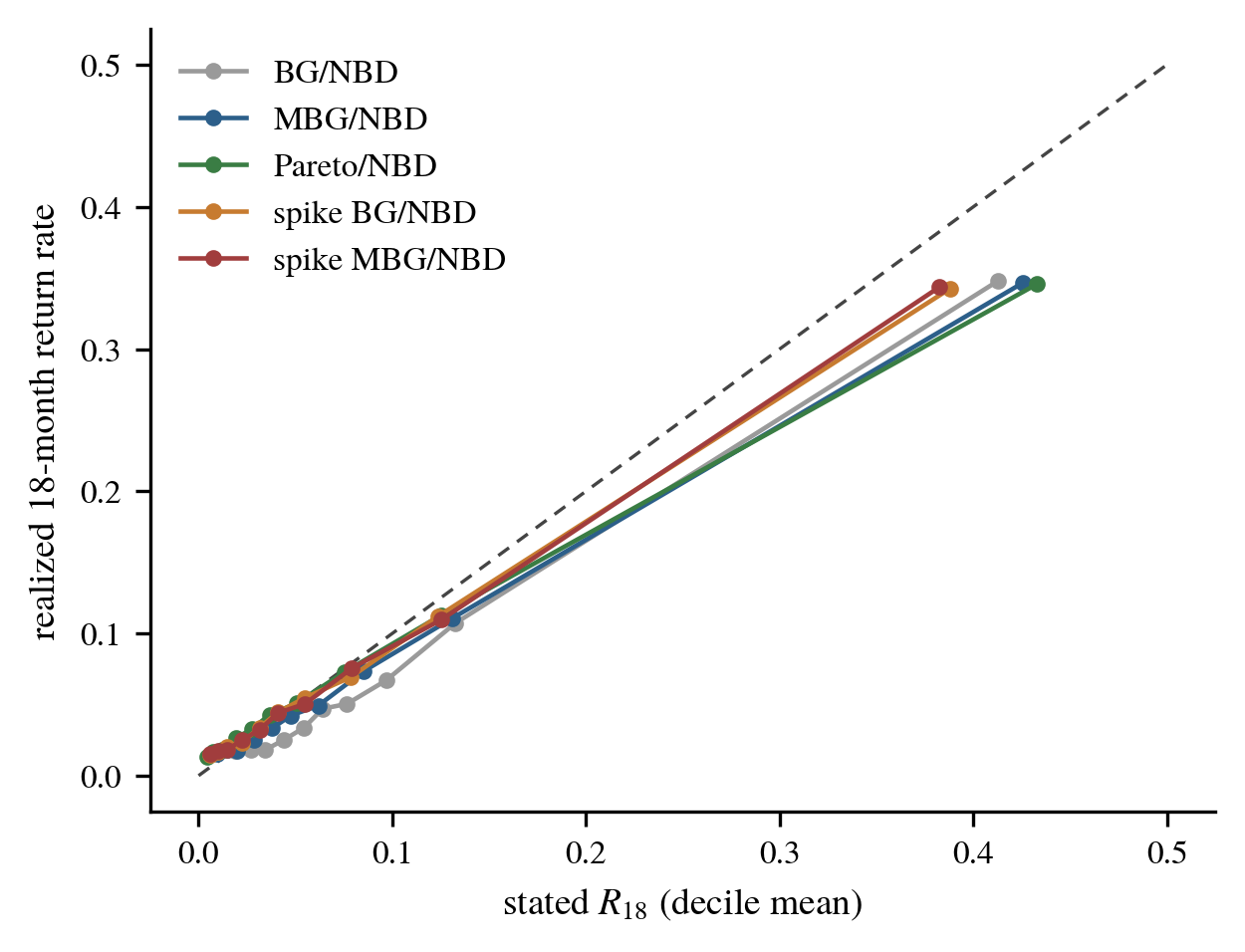}
\caption{The same fitted models' own 18-month forecasts: two outputs of one fitted object.}
\end{figure}

\hypertarget{the-dial}{%
\subsubsection{4.3 The dial}\label{the-dial}}

The estimation-consistent comparison at the same vintage assembles six variants: unpenalized maximum likelihood for five structures (BG/NBD, MBG/NBD, hard-core spike, Pareto/NBD, and the periodic death opportunity model of Jerath, Fader, and Hardie 2011, its death-opportunity period profiled from monthly to annual values) plus the MBG/NBD under the software-default ridge penalty (\(\rho = 0.01\)), which is what the production system runs. \(N_{18}\) spans only 2,433 to 3,047 against 2,319 realized returners; AUC is 0.79 to 0.81 throughout. \(N_\infty\) spans 3,654 to 27,734. The PDO deserves note because it severs the mechanical link between buying and dying that the beta-geometric family induces: it lands its count at 8,634, a fourth materially distinct point inside the interval, while its \(N_{18}\) of 2,634 sits inside the same narrow auditable band as everything else. Within the MBG/NBD alone the penalty moves the count from 3,660 to 5,211; the spike share is unidentified against the MBG/NBD's beta (their unpenalized fits coincide); the BG/NBD's inflation is structural and no penalty removes it. Everything the data can grade sits in a narrow band; the reported number spans a factor of 7.6, the balance supplied by conventions.

\begin{figure}
\centering
\includegraphics{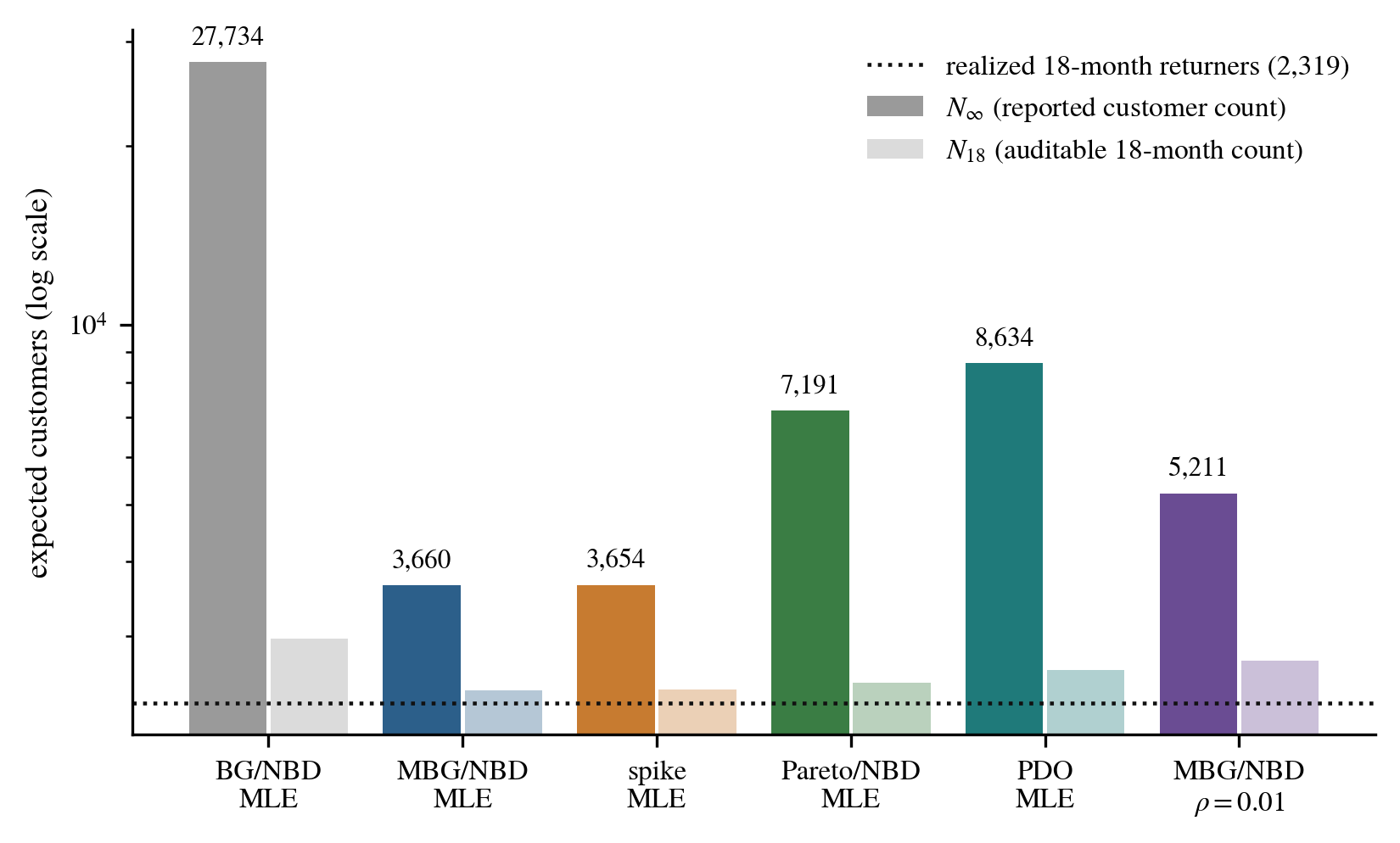}
\caption{The dial: estimation conventions set the reported count; auditable quantities barely move (2024 vintage).}
\end{figure}

\hypertarget{the-five-year-audit}{%
\subsubsection{4.4 The five-year audit}\label{the-five-year-audit}}

At the June 2021 vintage (2,967 consumer customers, 60 months of subsequent data), realized cumulative returners reach 313 by month 18 and 419 by month 60, still rising. Against that curve, the BG/NBD's summed aliveness of 2,711 is absurdly high; the unpenalized maximum-likelihood tail shared by the MBG/NBD (213) and the spike model (207) is crossed from below within thirty months, so the maximum-likelihood customer count is falsified by the firm's own subsequent data; the Pareto/NBD's 876 remains unfalsified, as does the PDO's 1,089, although the PDO's near-term trajectory under-predicts (213 predicted eighteen-month returners against 313 realized); and the penalized production fit's 474 tracks the realized curve closely at every horizon, an accuracy Section 4.8 shows to be an accident of where the penalty landed on the ridge. The identified interval \([L_v, U_v]\) for this vintage's \(N_\infty\) is bounded below by 419 and rising, with the maximum-likelihood point refuted and 2,711 excluded; no fitting criterion selects within what the record shows.

\begin{figure}
\centering
\includegraphics{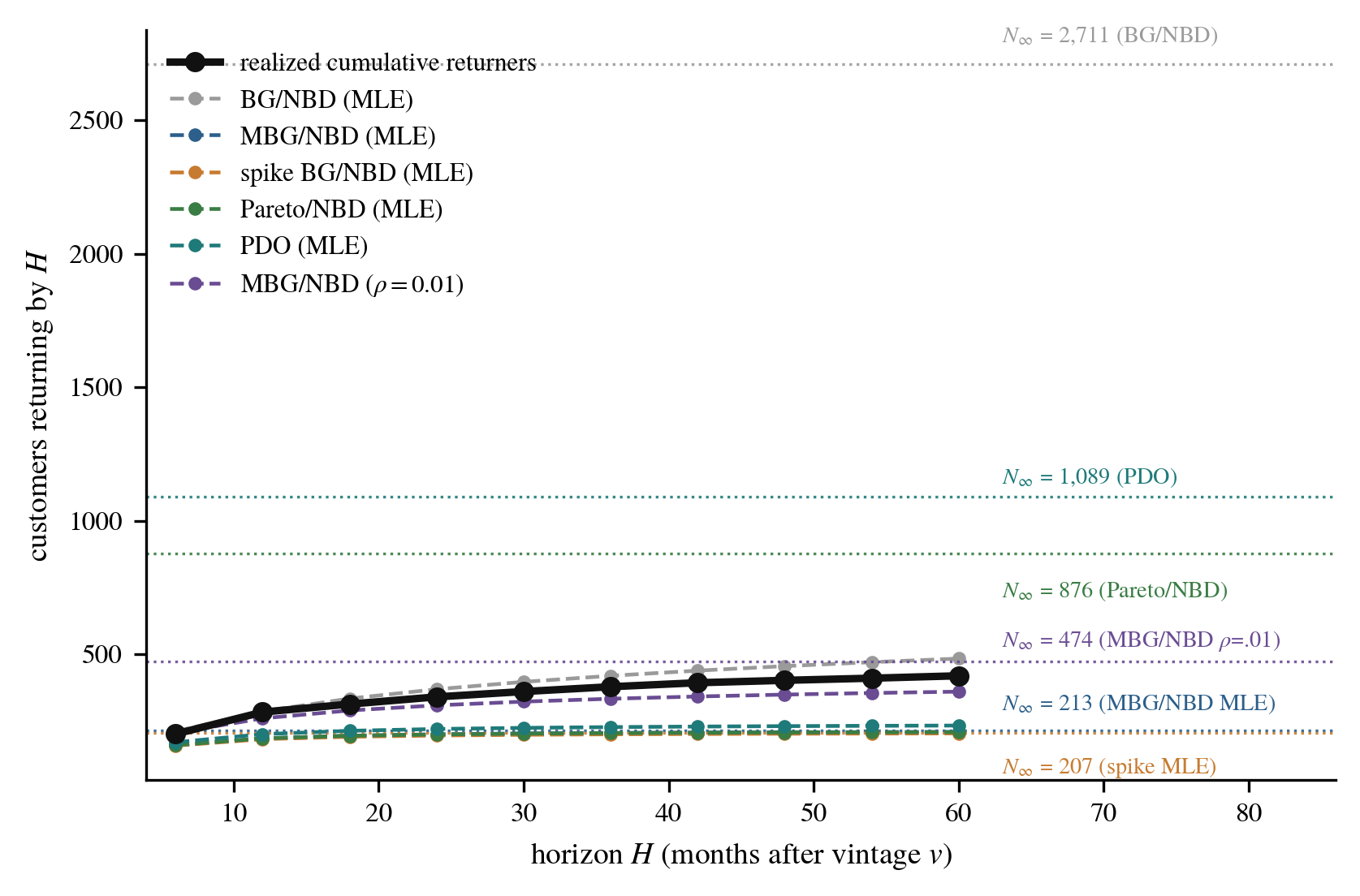}
\caption{The five-year audit: realized cumulative returners against each variant's trajectory and asymptote (2021 vintage).}
\end{figure}

\hypertarget{drift-and-the-impostor}{%
\subsubsection{4.5 Drift and the impostor}\label{drift-and-the-impostor}}

Native-forecast bias rises with vintage recency for every specification in both segments (consumer MBG/NBD: 1.00, 1.10, 1.18 across the 2022 through 2024 vintages; institutional variants: parity to 1.15-1.21), and even the aggregate order forecast drifts, so the movement is environment and composition rather than structure; the cohort triangle localizes the recent component to entering-cohort levels against stable development shapes, consistent with the firm's expansion into harder-to-reach segments of demand. The validation consequence is the impostor result: at the 2024 cell the best-calibrated eighteen-month forecasts belong to the spike and Pareto/NBD variants (1.05x and 1.08x against the MBG/NBD's 1.18x), the same spike variant whose tail the five-year audit falsified outright; its structural harshness cancels the contemporaneous softening. A validation exercise confined to the field's conventional single cell would select the refuted model.

\begin{figure}
\centering
\includegraphics{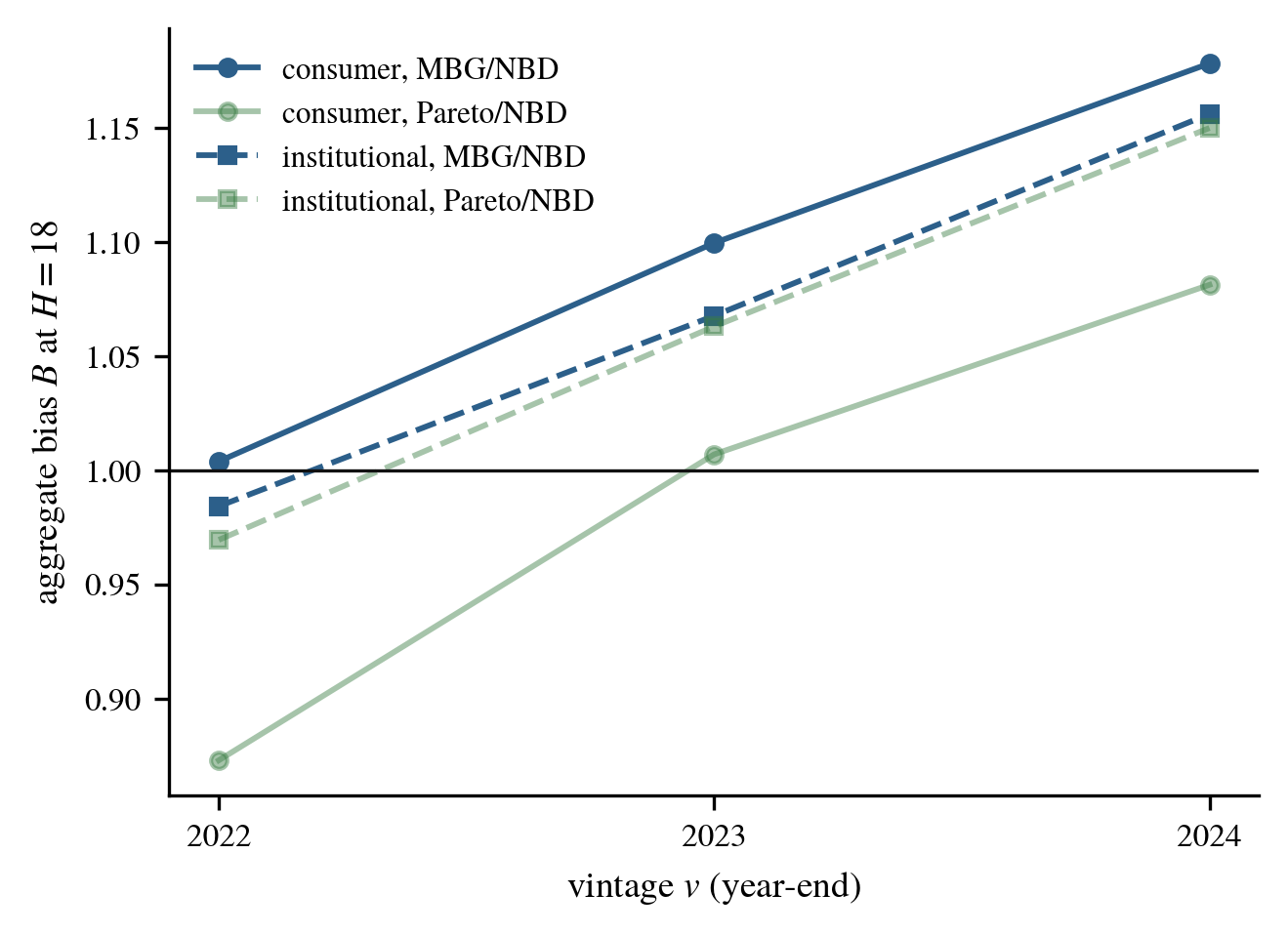}
\caption{Drift: native-forecast bias by vintage, both segments.}
\end{figure}

\hypertarget{cdnow-replication}{%
\subsubsection{4.6 CDNOW replication}\label{cdnow-replication}}

The identification patterns replicate on the canonical dataset of this literature. We use the full CDNOW master cohort: all 23,570 customers who made their first purchase at CDNOW in the first quarter of 1997, observed through June 1998 (69,659 transactions), as documented in Fader and Hardie's notes on the master data set; the 2,357-customer file familiar from Fader, Hardie, and Lee's published analyses is a 1/10 systematic sample of this cohort, and we use the full cohort for power. Under the standard 39-week calibration and 39-week holdout split there are 7,058 realized holdout returners. The dial: alive counts span 8,446 (spike, maximum likelihood) to 19,981 (BG/NBD, penalized), a 2.4x spread, on AUC flat at 0.76 to 0.77 and native 39-week forecast sums confined to 5,396 to 6,940. The penalty moves the MBG/NBD count 22 percent (8,484 to 10,350). The spike collapses into the smooth mixture exactly as in the field data (estimated share 0.45, counts within half a percent of the MBG/NBD's). The PDO is interior on both dimensions (13,215 alive against 7,058 realized returners, 1.87x, with a native forecast at 0.87x). The category error: summed aliveness overshoots realized returners by 1.20x to 2.83x across variants while the native forecasts sit at 0.81x to 0.98x, the models' familiar mild under-forecast of CDNOW's holdout. Magnitudes differ from the field panel, as they should (CDNOW's shorter window and different one-and-done share), but every qualitative claim, the flat-observables-divergent-counts dial, the penalty's leverage, the spike unidentifiability, and the \(A\)-versus-\(R_H\) wedge, transfers to the dataset on which this literature was built.

\begin{figure}
\centering
\includegraphics{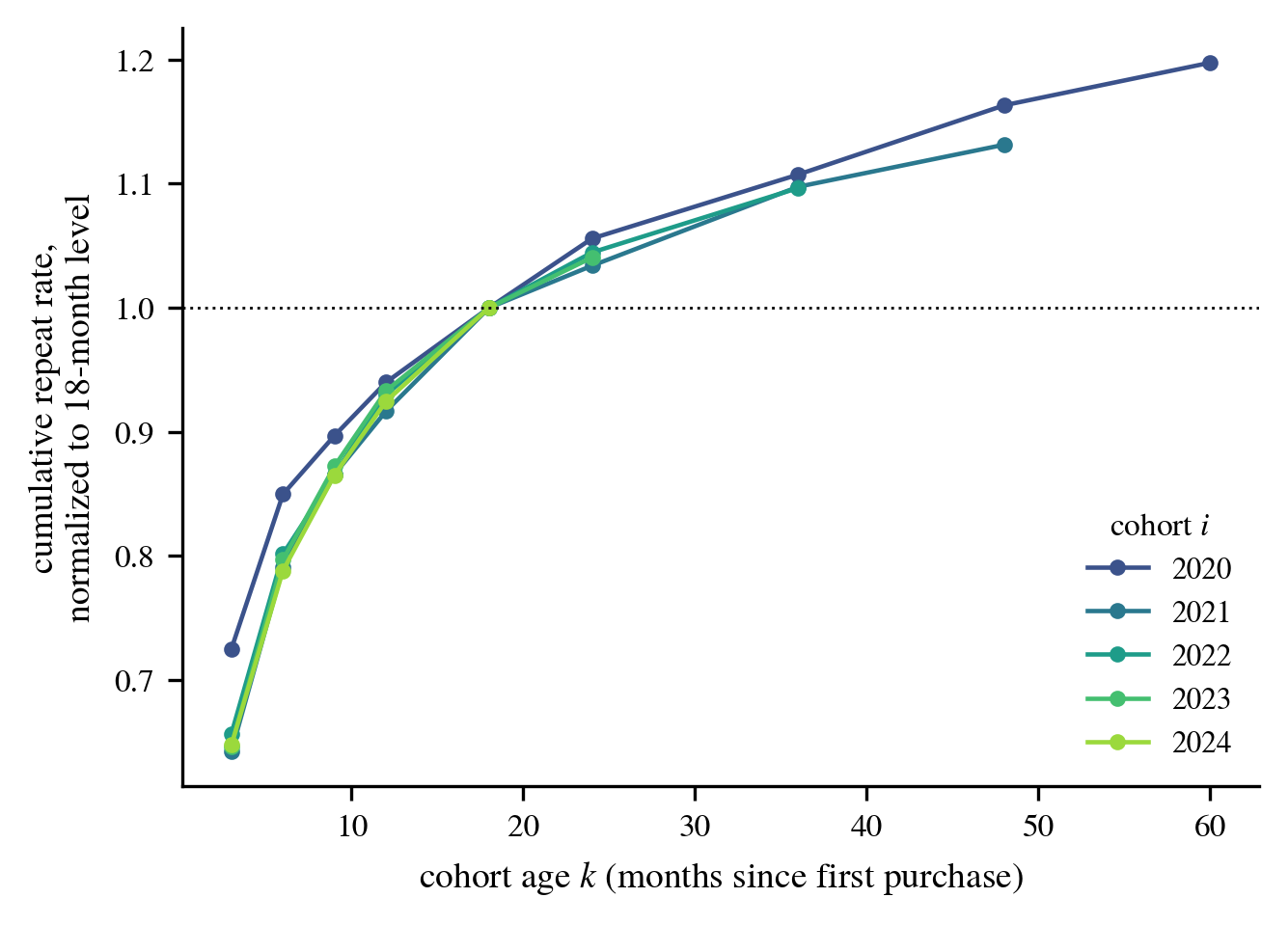}
\caption{Stable shape: cohort maturation curves normalized to the 18-month level (consumer regime).}
\end{figure}

\begin{figure}
\centering
\includegraphics{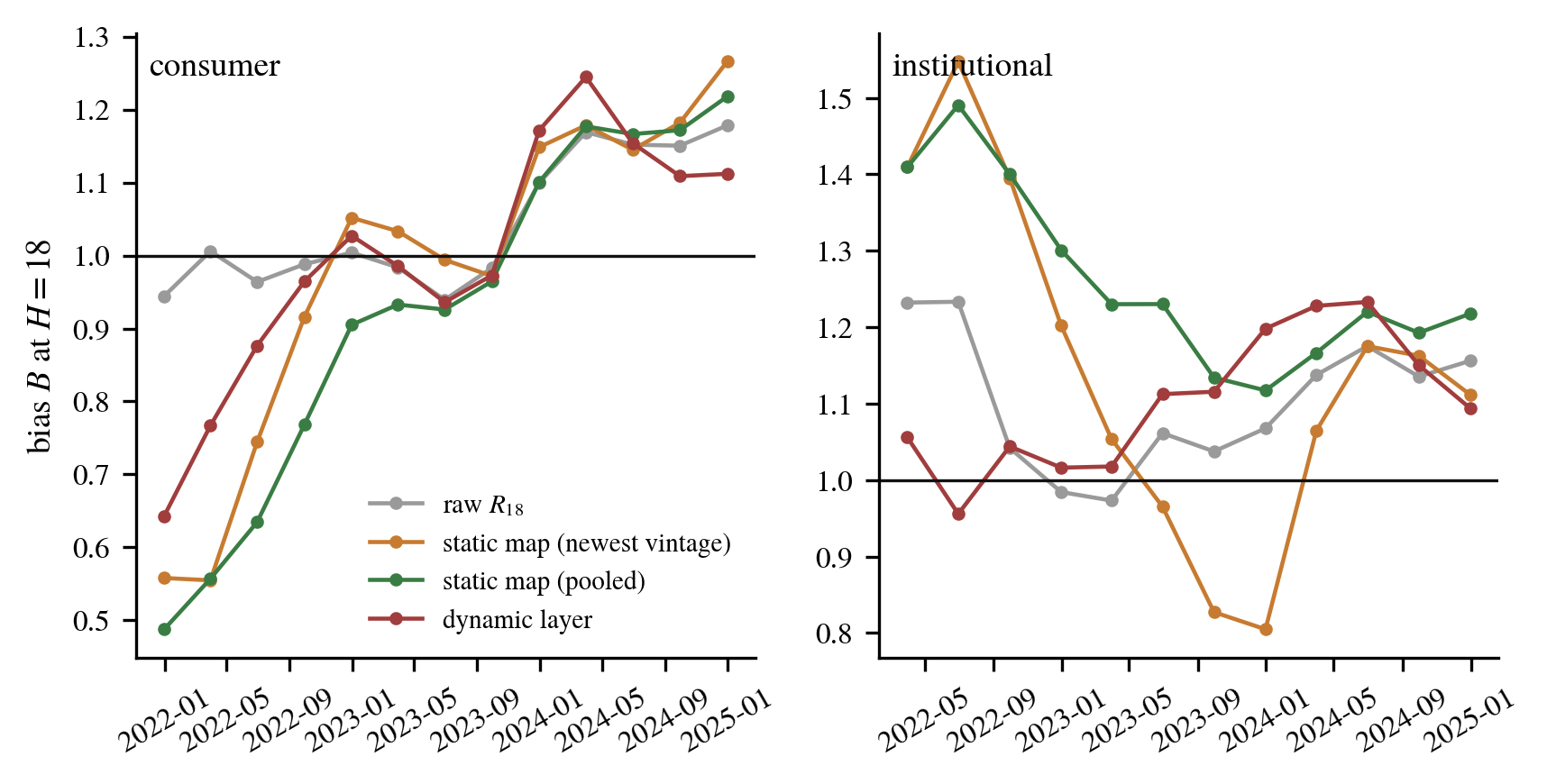}
\caption{Thirteen-quarter out-of-time backtest: aggregate bias \(B\) at \(H = 18\) by scoring vintage \(v\), for the raw forecast, static maps, and the dynamic layer; consumer and institutional regimes.}
\end{figure}

\begin{figure}
\centering
\includegraphics{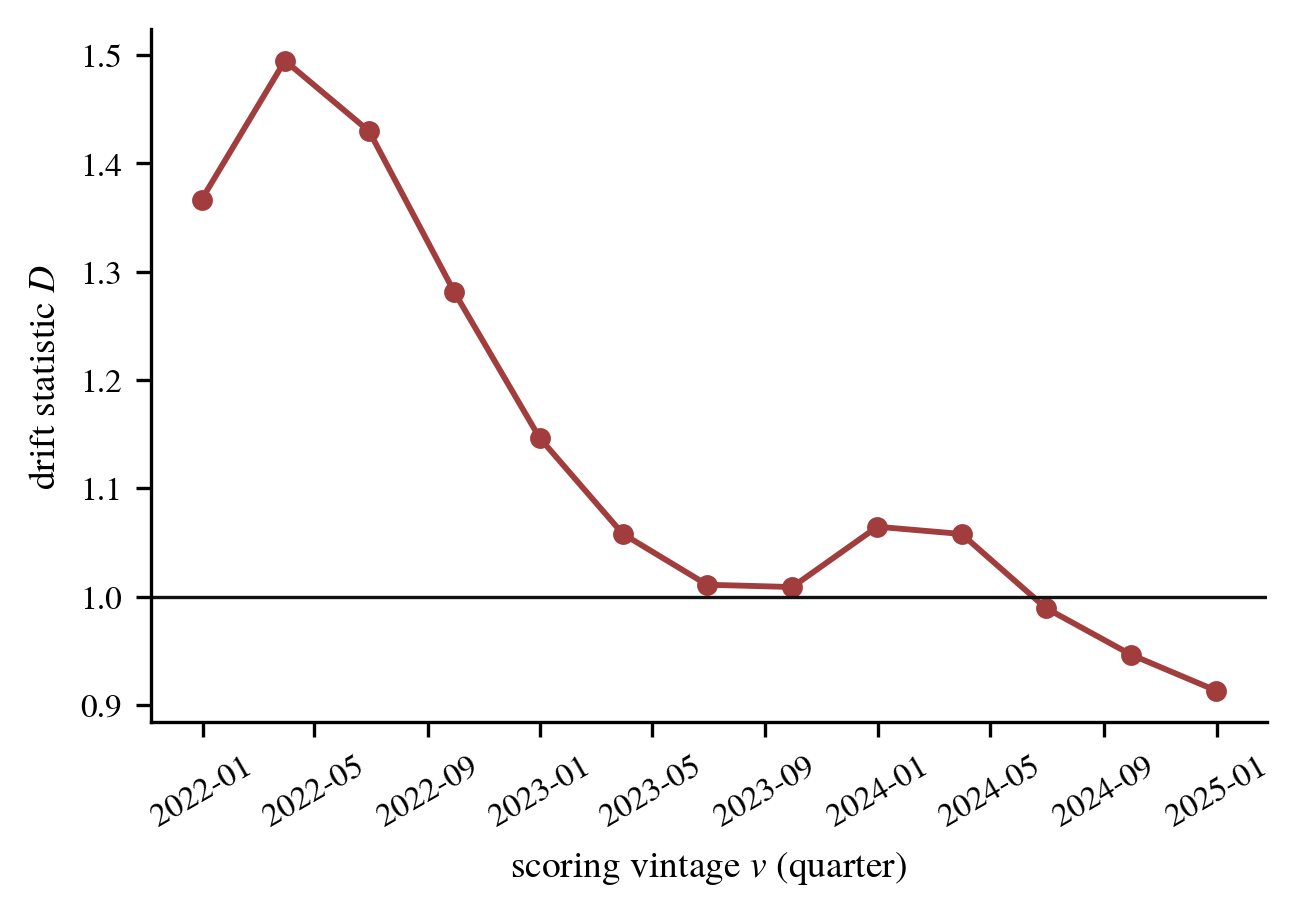}
\caption{The drift alarm: the drift statistic \(D\) by scoring vintage \(v\) (consumer regime). \(D = 1\) when the current regime matches the training regime; the mid-2024 crossing precedes visibility in full-horizon validation by two quarters.}
\end{figure}

\hypertarget{taking-fixes-taken-together}{%
\subsubsection{4.7 Taking fixes, taken together}\label{taking-fixes-taken-together}}

Figure 9 brings together the remedies. It pools every out-of-time forecast the dynamic layer produced across the backtest quarters (over 200,000 forecast-outcome pairs in the consumer regime, spanning the post-pandemic recovery, the stable middle, and the recent drift) and grades them at three horizons. The calibrated twelve-month forecast carries an aggregate bias of 1.06x and an expected calibration error of 0.004; eighteen months, 1.05x and 0.007; twenty-four months, 1.00x and 0.018. Behind each curve, in grey, sits the same customers' summed aliveness read as activity, the number a standard dashboard would have displayed. The dead-reckoned position drifts by a factor of two or more, while the position with fixes taken sits on the diagonal at every horizon a firm can audit, through regime changes, using only information available at each forecast date. The institutional segment, with two orders of magnitude fewer units, lands at 1.09x to 1.14x with wider scatter, which is what honest small-sample calibration looks like. Nothing in this figure requires abandoning the installed model; it is the same MBG/NBD everywhere, asked the answerable question and kept on course.

\begin{figure}
\centering
\includegraphics{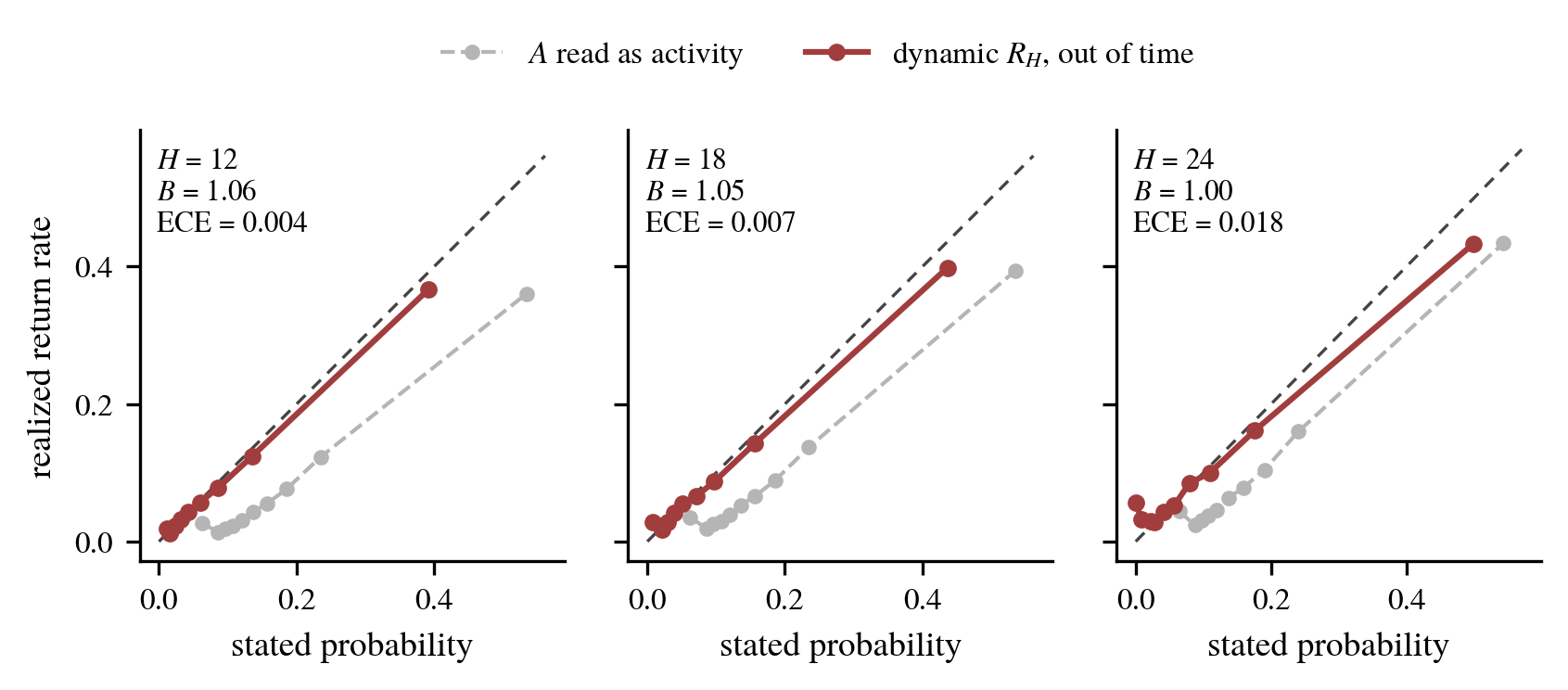}
\caption{The payoff: out-of-time reliability of the dynamically calibrated \(R_H\) at \(H\) = 12, 18, and 24 months, pooled across all backtest vintages (consumer regime; over 200,000 forecast-outcome pairs). Grey: the same customers' aliveness probability \(A\) read as activity, the dashboard default. Panel annotations give aggregate bias \(B\) and expected calibration error.}
\end{figure}

\hypertarget{mechanism-simulations-with-known-truth}{%
\subsubsection{4.8 Mechanism: simulations with known truth}\label{mechanism-simulations-with-known-truth}}

Three worlds, 30,000 customers each, isolate what the field data confound. In the smooth-truth world (an MBG/NBD at the field parameters), the unpenalized fitted model recovers its generating parameters and the true alive count (2,573 predicted, 2,584 true), and the posterior evaluated at true parameters is calibrated against true aliveness decile by decile: the audit machinery passes when nothing is wrong, so field rejections are not artifacts. In the spike-truth world (a genuine hard-core segment of 0.6), the spike fit recovers the share exactly while the smooth model absorbs it at a log-likelihood cost of 0.003 per observation; in the smooth world the free spike collapses to zero; the profile likelihood over the spike share is nearly flat, and the same pair of fitted objects is consistent with materially different truths. Refitting the smooth world at \(\rho = 0.01\) moves the alive count 53 percent above known truth while leaving every observable forecast unchanged to within rounding: the penalty buys its shrinkage along the one direction the likelihood does not price, which is the count. The drift world (a 30 percent rate cut partway through) reproduces the grid signatures of the field data, with bias moving all specifications together in calendar time and re-equilibrating after refits, confirming the decomposition logic the audit relies on. The worlds are deliberately minimal (stationary mixtures, one level shift, no covariates); they establish mechanism, not magnitude.

\hypertarget{managerial-implications}{%
\subsection{5. Managerial Implications}\label{managerial-implications}}

The operating system that follows has three instruments matched to three questions. Detection: the reliability diagram on the newest fully observed vintage is the standing health check, and the dynamic layer's drift statistic \(D\) is the between-audit alarm. Decomposition: when the alarm fires, the cohort triangle separates the world from the mix, since a macro shift moves every cohort off its own development curve simultaneously while a composition shift shows up as entering-cohort level changes against stable shapes and normally developing mature cohorts; our 2024 entering cohort reads as mix, the pandemic cohorts as macro, and the two can co-occur and cancel, which is why the decomposition belongs in the standing audit rather than in retrospective storytelling. Attribution: the firm's own interventions have no observational answer, and the point is about to become concrete, because the firm is preparing a reactivation program aimed at exactly the dormant customers where levels are least trustworthy. A winback rule is a breakeven computation on the level, and an unrandomized rollout destroys the audit that justified it: assignment by score removes the variation identification needs, bakes the treatment effect into the next recalibration, and silently changes the score's meaning from organic return propensity to return-under-current-treatment. The repair is one design decision made before the first contact: a randomized holdout (15 to 20 percent, stratified by score decile, or a staggered rollout with randomized batch order), after which the control stream carries the audit, the treated-minus-control gap by decile estimates incremental effect, and the dashboard grows from one dial to three: environment, mix, efficacy. We state the boundary plainly: a calibrated return probability identifies who is likely to return, not who returns because of intervention, and the two can diverge to the point of futility (Ascarza 2018), score-triggered retention interventions can backfire outright in the field (Ascarza, Iyengar, and Schleicher 2016), and fresh short-horizon signals can only partially repair long-horizon targeting (Huang and Ascarza 2024); the experiment identifies the second, and nothing in this paper's machinery does. Beyond the dashboard, the same rules are disclosure-grade. Customer-based corporate valuation (McCarthy and Fader 2018) is the important antecedent here: it takes customer projections seriously enough to build enterprise value on them and disciplines those projections against reported aggregates, and our interval standard extends exactly that discipline to the customer count itself. Valuations of noncontractual firms capitalize projections built on this machinery, an acquirer and a seller can currently both be right about ``how many customers'' to within a factor of several, and the interval-plus-auditable-horizon standard is what verification looks like.

The content of this section compresses into questions that any executive can ask of a customer dashboard, in the spirit of using questions as a summary of the framework. At what horizon has this number been audited, and when was the most recent fully observed vintage scored? What does the drift statistic \(D\) read this quarter, and who is notified when it moves? Which decisions in this firm consume the number as a level rather than as a ranking? Is anything the score triggers still unrandomized? And if the answer to the first question is that the number has never been audited, then the number is a reckoning, not a fix, and should be treated accordingly.

\hypertarget{concluding-remarks}{%
\subsection{6. Concluding Remarks}\label{concluding-remarks}}

The work in this paper began with a dashboard number that failed a backtest and ended somewhere different: the number was not wrong so much as unanswerable, and the machinery for saying so precisely turned out to be the useful contribution. We return to the three questions that motivated the research. First, what does the reported customer count mean, and can a finite record discipline it? It is the infinite-horizon limit of an auditable forecast family, and the record bounds it without determining it: the count is partially identified, and estimation conventions select the point. Second, how large are the consequences? A factor of 7.6 across observationally interchangeable specifications on our panel and 2.4x on CDNOW, 42 percent from a software default alone, and a category error that accounts for most of what practice would call miscalibration. Third, what should firms report instead? The calibrated horizon count \(N_H\) at a stated horizon, maintained by the dynamic layer, monitored by the drift statistic \(D\), and accompanied, when a total count must be stated, by the identified interval rather than a point.

We believe the contributions are as follows. Conceptually, we identify P(alive) as the infinite-horizon limit of an observable forecast family (exact in the BG family, an upper bound for the Pareto/NBD) and show the implied customer count is partially identified, bounded below by an accumulating observable and selected inside the bounds by conventions the data cannot audit. Empirically, we measure the consequences in an operating organization and replicate them on the canonical public dataset: a count spread of 7.6x (2.4x on CDNOW) across observationally interchangeable specifications, a 42 percent (22 percent) count movement from a default regularization constant, falsification of the maximum-likelihood count by five years of patience, a category error that accounts for most of what practice would call miscalibration, and a demonstration that the field's single-split validation design can select a structurally refuted model whose bias cancels drift. Methodologically, we assemble standard tools into the audit and repair that follow (the vintage-by-horizon grid, the cohort-credibility calibration layer with its drift alarm, validated head-to-head against cheap structural alternatives) and into a governance architecture that assigns detection to the grid, decomposition to the cohort triangle, and attribution to randomized rollout. The practitioner summary is: navigate by dead reckoning if you must, but take fixes, correct course, and never log the extrapolated position as an observed one. Report \(N_H\); monitor \(D\); state \(N_\infty\) as the interval \([L_v, U_v]\) if it must be stated at all; audit the \((v, H)\) grid, not the cell; and randomize anything the score triggers.

\textbf{Limitations and the road ahead.} The analysis is limited in five ways. First, the empirical results come from one firm plus one public benchmark, and generality is claimed only in terms of the mechanism: the identification results are model-class properties verified in simulation and replicated on CDNOW; the magnitudes are idiosyncratic. Second, the stable-shape regularity behind the dynamic layer is an empirical conjecture about noncontractual purchasing whose actuarial ancestor has held across a century of insurance lines, which is encouragement but not really evidence. Third, inference is the least mature component: thirteen quarters of overlapping windows give wide, overlapping intervals for the method comparison, the drift-regime superiority of the dynamic layer rests on four quarters and one drift episode, and the deep-horizon audit consumes calendar time by nature. Fourth, we did not run the horse race against discriminative machine learning, because our claims concern auditing structurally interpreted models rather than ranking predictors; the audit is model-agnostic and would grade a boosted challenger identically, and the challenger reports no alive count at all (recurrent-network models of customer-base activity indeed report none; Valendin et al.~2022), which under this paper's findings may be a feature. Fifth, covariates are deliberately absent from both model and layer, to keep the contribution legible, with the extensions mapped (cohort-specific tilts, changepoint and time-varying structures); this includes seasonality, whose outcome-side effects the whole-year and six-quarter horizons largely integrate out but whose fitting-side effects remain unmodeled; the Gaussian-process machinery of Dew and Ansari (2018) and Dew, Ansari, and Li (2020) is the natural structural route for calendar and seasonal covariates, and the audit developed here is how any such extension would demonstrate its gains. Efficacy is the sequel: the reactivation experiment will estimate what no calibration can. Revenue and profit are out of scope: all results operate on purchase counts, which keeps the product-versus-factorization logic clean, and value-weighting via spend submodels raises the same questions one level up, since the spend distribution's tail is another weakly disciplined extrapolation.
\newpage

\hypertarget{references}{%
\subsection{References}\label{references}}

Abe M (2009) ``Counting your customers'' one by one: A hierarchical Bayes extension to the Pareto/NBD model. \emph{Marketing Science} 28(3):541-553.

Andrews DWK (1988) Chi-square diagnostic tests for econometric models: Introduction and applications. \emph{Journal of Econometrics} 37(1):135-156.

Ascarza E (2018) Retention futility: Targeting high-risk customers might be ineffective. \emph{Journal of Marketing Research} 55(1):80-98.

Ascarza E, Iyengar R, Schleicher M (2016) The perils of proactive churn prevention using plan recommendations: Evidence from a field experiment. \emph{Journal of Marketing Research} 53(1):46-60.

Ascarza E, Netzer O, Hardie BGS (2018) Some customers would rather leave without saying goodbye. \emph{Marketing Science} 37(1):54-77.

Bachmann P, Meierer M, Näf J (2021) The role of time-varying contextual factors in latent attrition models for customer base analysis. \emph{Marketing Science} 40(4):783-809.

Batislam EP, Denizel M, Filiztekin A (2007) Empirical validation and comparison of models for customer base analysis. \emph{International Journal of Research in Marketing} 24(3):201-209.

Bornhuetter RL, Ferguson RE (1972) The actuary and IBNR. \emph{Proceedings of the Casualty Actuarial Society} 59:181-195.

Brier GW (1950) Verification of forecasts expressed in terms of probability. \emph{Monthly Weather Review} 78(1):1-3.

Chamberlain BP, Cardoso A, Liu CHB, Pagliari R, Deisenroth MP (2017) Customer lifetime value prediction using embeddings. \emph{Proceedings of the 23rd ACM SIGKDD International Conference on Knowledge Discovery and Data Mining} (ACM, New York), 1753-1762.

Cox DR (1958) Two further applications of a model for binary regression. \emph{Biometrika} 45(3-4):562-565.

Davis SE, Lasko TA, Chen G, Siew ED, Matheny ME (2017) Calibration drift in regression and machine learning models for acute kidney injury. \emph{Journal of the American Medical Informatics Association} 24(6):1052-1061.

Demler OV, Paynter NP, Cook NR (2015) Tests of calibration and goodness-of-fit in the survival setting. \emph{Statistics in Medicine} 34(10):1659-1680.

Dew R, Ansari A (2018) Bayesian nonparametric customer base analysis with model-based visualizations. \emph{Marketing Science} 37(2):216-235.

Dew R, Ansari A, Li Y (2020) Modeling dynamic heterogeneity using Gaussian processes. \emph{Journal of Marketing Research} 57(1):55-77.

Dimitriadis T, Fader PS, Hardie BGS, Shang J (2010) Customer-base analysis in a discrete-time noncontractual setting. \emph{Marketing Science} 29(6):1086-1108.

Gneiting T, Jordan AI (2021) Stable reliability diagrams for probabilistic classifiers. \emph{Proceedings of the National Academy of Sciences} 118(8):e2016191118.

Fader PS, Hardie BGS (2001) Notes on the CDNOW master data set. Technical note, http://brucehardie.com/notes/026/.

Fader PS, Hardie BGS (2014) The Pareto/NBD is not a ``lost-for-good'' model. Technical note, http://brucehardie.com/notes/031/.

Fader PS, Hardie BGS, Lee KL (2005) ``Counting your customers'' the easy way: An alternative to the Pareto/NBD model. \emph{Marketing Science} 24(2):275-284.

Gneiting T, Raftery AE (2007) Strictly proper scoring rules, prediction, and estimation. \emph{Journal of the American Statistical Association} 102(477):359-378.

Gopalakrishnan A, Bradlow ET, Fader PS (2017) A cross-cohort changepoint model for customer-base analysis. \emph{Marketing Science} 36(2):195-213.

Graf E, Schmoor C, Sauerbrei W, Schumacher M (1999) Assessment and comparison of prognostic classification schemes for survival data. \emph{Statistics in Medicine} 18(17-18):2529-2545.

Guo C, Pleiss G, Sun Y, Weinberger KQ (2017) On calibration of modern neural networks. \emph{Proceedings of the 34th International Conference on Machine Learning}, PMLR 70:1321-1330.

Hoppe D, Wagner U (2007) Customer base analysis: The case for a central variant of the Betageometric/NBD model. \emph{Marketing: Journal of Research and Management} 3(2):75-90.

Huang TW, Ascarza E (2024) Doing more with less: Overcoming ineffective long-term targeting using short-term signals. \emph{Marketing Science} 43(4):863-884.

Hosmer DW, Lemeshow S (1980) Goodness of fit tests for the multiple logistic regression model. \emph{Communications in Statistics: Theory and Methods} 9(10):1043-1069.

Jerath K, Fader PS, Hardie BGS (2011) New perspectives on customer ``death'' using a generalization of the Pareto/NBD model. \emph{Marketing Science} 30(5):866-880.

Lemmens A, Gupta S (2020) Managing churn to maximize profits. \emph{Marketing Science} 39(5):956-973.

Manski CF (2003) \emph{Partial Identification of Probability Distributions}. Springer Series in Statistics (Springer, New York).

Martínez A, Schmuck C, Pereverzyev S Jr, Pirker C, Haltmeier M (2020) A machine learning framework for customer purchase prediction in the non-contractual setting. \emph{European Journal of Operational Research} 281(3):588-596.

McCarthy DM, Fader PS (2018) Customer-based corporate valuation for publicly traded noncontractual firms. \emph{Journal of Marketing Research} 55(5):617-635.

McCarthy DM, Oblander ES (2021) Scalable data fusion with selection correction: An application to customer base analysis. \emph{Marketing Science} 40(3):459-480.

Murphy AH (1973) A new vector partition of the probability score. \emph{Journal of Applied Meteorology} 12(4):595-600.

Neslin SA, Gupta S, Kamakura W, Lu J, Mason CH (2006) Defection detection: Measuring and understanding the predictive accuracy of customer churn models. \emph{Journal of Marketing Research} 43(2):204-211.

Padilla N, Ascarza E (2021) Overcoming the cold start problem of customer relationship management using a probabilistic machine learning approach. \emph{Journal of Marketing Research} 58(5):981-1006.

Platzer M, Reutterer T (2016) Ticking away the moments: Timing regularity helps to better predict customer activity. \emph{Marketing Science} 35(5):779-799.

Reutterer T, Platzer M, Schröder N (2021) Leveraging purchase regularity for predicting customer behavior the easy way. \emph{International Journal of Research in Marketing} 38(1):194-215.

Rust RT, Lemon KN, Zeithaml VA (2004) Return on marketing: Using customer equity to focus marketing strategy. \emph{Journal of Marketing} 68(1):109-127.

Schmittlein DC, Morrison DG, Colombo R (1987) Counting your customers: Who are they and what will they do next? \emph{Management Science} 33(1):1-24.

Spiegelhalter DJ (1986) Probabilistic prediction in patient management and clinical trials. \emph{Statistics in Medicine} 5(5):421-433.

Van Calster B, McLernon DJ, van Smeden M, Wynants L, Steyerberg EW (2019) Calibration: The Achilles heel of predictive analytics. \emph{BMC Medicine} 17:230.

Van Calster B, Nieboer D, Vergouwe Y, De Cock B, Pencina MJ, Steyerberg EW (2016) A calibration hierarchy for risk models was defined: From utopia to empirical data. \emph{Journal of Clinical Epidemiology} 74:167-176.

Valendin J, Reutterer T, Platzer M, Kalcher K (2022) Customer base analysis with recurrent neural networks. \emph{International Journal of Research in Marketing} 39(4):988-1018.

Wübben M, von Wangenheim F (2008) Instant customer base analysis: Managerial heuristics often ``get it right.'' \emph{Journal of Marketing} 72(3):82-93.

Wünderlich R (2015) Hierarchical Bayesian models for predicting purchase behavior in noncontractual settings based on individual and cross-sectional purchase patterns. Doctoral dissertation, Technische Universität München.

Zadrozny B, Elkan C (2002) Transforming classifier scores into accurate multiclass probability estimates. \emph{Proceedings of the 8th ACM SIGKDD International Conference on Knowledge Discovery and Data Mining} (ACM, New York), 694-699.

\newpage

\hypertarget{appendix}{%
\subsection{Appendix}\label{appendix}}

\hypertarget{a1.-partial-identification-of-the-customer-count}{%
\subsubsection{A1. Partial identification of the customer count}\label{a1.-partial-identification-of-the-customer-count}}

\textbf{Setup.} A cohort of \(n\) customers is observed from first purchase through vintage \(v\), and thereafter through \(v + W\), where \(W\) is the elapsed post-vintage observation window. Let \(a_j \in \{0, 1\}\) indicate that customer \(j\) is alive at \(v\), and let \(Y_j(W) \in \{0,1\}\) indicate at least one purchase in \((v, v+W]\). The estimand is the realized alive count \(N^* = \sum_j a_j\); a fitted model reports \(\hat{N}_\infty = \sum_j \hat{A}_j\) as its estimate of \(N^*\).

\textbf{Proposition (informal statement; BG family).} (i) \emph{Limit identity.} Because dropout occurs only at a purchase, a customer alive at \(v\) purchases again with probability one given unlimited time, so \(a_j = \lim_{W \to \infty} Y_j(W)\) almost surely, and correspondingly \(A = \lim_{H \to \infty} R_H\) and \(N_\infty = \lim_{H \to \infty} N_H\). (For the Pareto/NBD, death can precede the next purchase, so \(\lim_W Y_j(W) \leq a_j\) and \(A\) bounds every observable member from above; the bounds below hold a fortiori.) (ii) \emph{Accumulating lower bound.} \(L_v(W) = \sum_j Y_j(W) \leq N^*\) almost surely, is nondecreasing in \(W\), and under the BG family converges to \(N^*\): the count is point-identified at \(W = \infty\) and bounded below at every finite \(W\). (iii) \emph{Sharp nonparametric bounds.} Absent restrictions on the purchase-rate distribution, the identified set for \(N^*\) given data through \(v + W\) is \([L_v(W),\, n]\). Sketch: for any target value in the interval, augment the population with customers who are alive but have purchase rate \(\lambda \to 0^+\); such customers are observationally equivalent to dead ones on any finite window (their likelihood contribution converges to that of a dead customer), while each contributes one unit to \(N^*\). An alive-but-arbitrarily-slow customer and a dead customer differ only in the tail the window cannot reach. (iv) \emph{Convention dependence within a parametric family.} A parametric mixing family restricts the set, but only through its tail assumptions: when the log-likelihood is nearly flat along a direction \(u\) in parameter space that moves tail mass (the empirical situation documented in Sections 4.3 and 4.8, where profile log-likelihood differences along \(u\) are a fraction of a point per ten thousand customers), any tie-breaking device selects the reported \(\hat{N}_\infty\). Penalized estimation with ridge constant \(\rho\) solves \(\max_\theta \{ \ell(\theta) - \rho \|\theta\|^2 \}\); along \(u\) the first term is nearly constant, so the penalty term governs the location of the optimum in that direction, and \(\partial \hat{N}_\infty / \partial \rho \neq 0\) even as fitted values of every finite-horizon functional are unchanged to first order. Priors in Bayesian implementations play the identical role through the same mechanism. The vocabulary of partial identification is Manski's (2003); the application to latent-attrition counts is, to our knowledge, new.

\textbf{Remarks.} The upper bound \(n\) is informative only jointly with a maintained assumption bounding purchase rates away from zero among the living (equivalently, an assumed saturation horizon); firms may find such an assumption reasonable (a customer who will not return within a decade is dead for every managerial purpose), and adopting one converts the interval to \([L_v(W),\, L_v(W)/q(W)]\) where \(q(W)\) is the assumed minimum probability that a living customer returns within \(W\). This is a disciplined way to report the count if a count must be reported; the assumption, not the data, supplies the upper bound, and should be stated. Human and firm lifespans bound ``eventually'' in practice, which is why we describe the count as unanswerable on managerial timescales rather than metaphysically: Section 4.4's five-year window narrows the interval materially and falsifies particular conventions, and that is exactly the process the proposition licenses.

\hypertarget{a2.-the-audit-protocol}{%
\subsubsection{A2. The audit protocol}\label{a2.-the-audit-protocol}}

For each segment and each fully observed cell \((v, H)\) (vintages quarterly; \(v + H\) before the data's end): fit the model on transactions through \(v\) only; compute native \(R_H\) for every customer in the base at \(v\); record outcomes on \((v, v+H]\). Report per cell: the CORP reliability diagram (isotonic regression of outcomes on forecasts) with consistency bands, aggregate bias \(B\), the Brier score with its Murphy (1973) decomposition, and AUC. Summarize per model across the grid by mean absolute log bias and by the deep-horizon check of predicted trajectories and \(\hat{N}_\infty\) asymptotes against cumulative realized returners. Make interval statements at the vintage level: a moving-block bootstrap over the quarterly cell-bias series (block length four quarters, respecting the serial dependence induced by overlapping windows), with within-cell customer bootstraps reported only as conditional-on-window statements. Never grade a fitted correction on its own training vintage. Where partially observed cells must be used, replace the Brier score with its inverse-probability-of-censoring-weighted version and grouped calibration tests with the Greenwood-Nam-D'Agostino form.

\hypertarget{a3.-the-dynamic-layer-and-the-penalty-mechanism}{%
\subsubsection{A3. The dynamic layer and the penalty mechanism}\label{a3.-the-dynamic-layer-and-the-penalty-mechanism}}

\textbf{Tilt estimator.} At vintage \(v\) with operating horizon \(H\): the shape \(g\) is the isotonic regression of outcomes on native \(R_H\) forecasts pooled over all fully observed training vintages \(\{v' : v' + H \leq v\}\). For short horizons \(m \in \{3, 6, 9, 12\}\) months, define the aggregate short-horizon bias \(b_m(v') = \sum_j R_{m,j}(v') / \sum_j Y_j(v', m)\). The current diagonal comprises the freshest partially observed vintages: \(b_m(v - m)\) for each \(m\), each fully observed by construction at scoring time. The drift statistic is the precision-weighted ratio of training-regime to current-regime short-horizon bias, \(D(v) = \sum_m w_m \bar{b}_m^{train} / \sum_m w_m b_m(v-m)\) with weights \(w_m\) equal to realized returner counts, and the deployed score is \(\min\{1, \max\{0, D(v) \cdot g(R_H)\}\}\). Ranking is preserved because \(g\) is monotone and \(D\) is a positive scalar. The backtest protocol and the head-to-head design (per-cohort and rolling-window re-estimation on the identical grid) follow Section 3.2, with all methods scored on identical outcome sets.

\textbf{Penalty mechanics.} Standard fitting software (the lifetimes package and its descendants; our production system) maximizes the mean log-likelihood minus \(\rho \sum_k \theta_k^2\), with \(\rho = 0.01\) a conventional default. The BG-family likelihood is nearly flat along the direction that trades tail mass of the dropout mixture against the purchase-rate distribution (Section 2.1); along that direction the penalty term is the only meaningfully varying component of the objective, so the penalized optimum slides to smaller parameter norms, which under the beta-gamma parameterization shifts posterior aliveness upward for the low-frequency mass. The observable consequence is the signature documented in the text: at \(\rho = 0.01\) versus \(\rho = 0\), \(\hat{N}_\infty\) moves 53 percent against known truth in simulation, 42 percent on the field panel, and 22 percent on CDNOW, while finite-horizon forecasts, likelihood, and AUC are unchanged to within rounding. Nothing is special about ridge penalties; any tie-breaker on a flat direction (a prior, an initialization, a convergence tolerance) has the same license.

\hypertarget{a4.-practice-survey-how-palive-is-consumed}{%
\subsubsection{A4. Practice survey: how P(alive) is consumed}\label{a4.-practice-survey-how-palive-is-consumed}}

Sources verified by direct retrieval on July 3, 2026; quotes verbatim; documentation versions as accessed on that date (the lifetimes quickstart at v0.11.2 documentation, PyMC-Marketing at the stable documentation build). The survey supports four claims: exposure of per-customer P(alive) is universal in the major implementations; thresholding and churn-dashboard use are routine in practitioner materials; horizon-consumption occurs inside CLV formulas and inside official documentation language; explicit summation into active-customer counts appears mainly in valuation-adjacent framing rather than tutorials. We found no source in the software or practitioner literature that warns against reading P(alive) as a horizon-return probability; the warning exists in the academic literature (Fader, Hardie, and Shang 2010) and did not propagate to the tooling.

\begin{longtable}[]{@{}lll@{}}
\toprule
\begin{minipage}[b]{0.30\columnwidth}\raggedright
Source (type)\strut
\end{minipage} & \begin{minipage}[b]{0.30\columnwidth}\raggedright
Use pattern\strut
\end{minipage} & \begin{minipage}[b]{0.30\columnwidth}\raggedright
Verbatim quote (abridged)\strut
\end{minipage}\tabularnewline
\midrule
\endhead
\begin{minipage}[t]{0.30\columnwidth}\raggedright
lifetimes Quickstart (docs)\strut
\end{minipage} & \begin{minipage}[t]{0.30\columnwidth}\raggedright
exposes; frames as ``still a customer''\strut
\end{minipage} & \begin{minipage}[t]{0.30\columnwidth}\raggedright
``What are the chances they are still `alive'? \ldots{} we can calculate their historical probability of being alive''\strut
\end{minipage}\tabularnewline
\begin{minipage}[t]{0.30\columnwidth}\raggedright
lifetimes API (docs)\strut
\end{minipage} & \begin{minipage}[t]{0.30\columnwidth}\raggedright
exposes per-customer quantity\strut
\end{minipage} & \begin{minipage}[t]{0.30\columnwidth}\raggedright
``Compute the probability that a customer with history (frequency, recency, T) is currently alive''\strut
\end{minipage}\tabularnewline
\begin{minipage}[t]{0.30\columnwidth}\raggedright
R BTYD, bgnbd.PAlive (docs)\strut
\end{minipage} & \begin{minipage}[t]{0.30\columnwidth}\raggedright
exposes; plots distribution across customers\strut
\end{minipage} & \begin{minipage}[t]{0.30\columnwidth}\raggedright
``the probability that they are still alive at the end of the calibration period''\strut
\end{minipage}\tabularnewline
\begin{minipage}[t]{0.30\columnwidth}\raggedright
R CLVTools vignette (docs)\strut
\end{minipage} & \begin{minipage}[t]{0.30\columnwidth}\raggedright
PAlive is a default predict() column\strut
\end{minipage} & \begin{minipage}[t]{0.30\columnwidth}\raggedright
``\,`probability of a customer being alive' (PAlive) at the end of the estimation period \ldots{} for every customer''\strut
\end{minipage}\tabularnewline
\begin{minipage}[t]{0.30\columnwidth}\raggedright
PyMC-Marketing Quickstart (docs)\strut
\end{minipage} & \begin{minipage}[t]{0.30\columnwidth}\raggedright
glosses P(alive) as horizon return\strut
\end{minipage} & \begin{minipage}[t]{0.30\columnwidth}\raggedright
plot titled ``Probability Customer 1516 will purchase again''; ``the model has low confidence that the customer will ever return''\strut
\end{minipage}\tabularnewline
\begin{minipage}[t]{0.30\columnwidth}\raggedright
Onnen, Towards Data Science 2021 (tutorial)\strut
\end{minipage} & \begin{minipage}[t]{0.30\columnwidth}\raggedright
dashboard column; churn risk = 1 minus P(alive); 0.9 filter\strut
\end{minipage} & \begin{minipage}[t]{0.30\columnwidth}\raggedright
``Its complement (1 - p) is equivalent to the customer's churn risk''\strut
\end{minipage}\tabularnewline
\begin{minipage}[t]{0.30\columnwidth}\raggedright
Blendo (vendor blog)\strut
\end{minipage} & \begin{minipage}[t]{0.30\columnwidth}\raggedright
thresholds at 0.5\strut
\end{minipage} & \begin{minipage}[t]{0.30\columnwidth}\raggedright
``customers with a probability \ldots{} less that 0.5 can be considered to be `dangerous'\,''\strut
\end{minipage}\tabularnewline
\begin{minipage}[t]{0.30\columnwidth}\raggedright
Windsor Circle (production vendor)\strut
\end{minipage} & \begin{minipage}[t]{0.30\columnwidth}\raggedright
ships 1 - P(alive) as churn score; validates against holdout purchases\strut
\end{minipage} & \begin{minipage}[t]{0.30\columnwidth}\raggedright
``the probability that this customer is dead to us grows \ldots{} That's the churn score''\strut
\end{minipage}\tabularnewline
\begin{minipage}[t]{0.30\columnwidth}\raggedright
Aliz.ai / Antonio (tutorial)\strut
\end{minipage} & \begin{minipage}[t]{0.30\columnwidth}\raggedright
horizon consumption in CLV formula\strut
\end{minipage} & \begin{minipage}[t]{0.30\columnwidth}\raggedright
assumes ``the probability-of-being-alive p remains unchanged in the next k periods''\strut
\end{minipage}\tabularnewline
\begin{minipage}[t]{0.30\columnwidth}\raggedright
Rietveld, bookdown (tutorial)\strut
\end{minipage} & \begin{minipage}[t]{0.30\columnwidth}\raggedright
glosses as future-purchase probability; buckets 0.25-0.8 as churners\strut
\end{minipage} & \begin{minipage}[t]{0.30\columnwidth}\raggedright
``Probability of being alive (which means customer who will purchase in the future)''\strut
\end{minipage}\tabularnewline
\begin{minipage}[t]{0.30\columnwidth}\raggedright
Munro, Towards Data Science 2024 (tutorial)\strut
\end{minipage} & \begin{minipage}[t]{0.30\columnwidth}\raggedright
constant multiplier over n transactions\strut
\end{minipage} & \begin{minipage}[t]{0.30\columnwidth}\raggedright
``alive probability and purchase value will stay fairly constant over the next n transactions''\strut
\end{minipage}\tabularnewline
\begin{minipage}[t]{0.30\columnwidth}\raggedright
Theta CBCV glossary (vendor)\strut
\end{minipage} & \begin{minipage}[t]{0.30\columnwidth}\raggedright
valuation framing around aliveness\strut
\end{minipage} & \begin{minipage}[t]{0.30\columnwidth}\raggedright
BTYD models explain ``how frequently customers make purchases while they are still `alive'\,''\strut
\end{minipage}\tabularnewline
\begin{minipage}[t]{0.30\columnwidth}\raggedright
Rust, Lemon, and Zeithaml 2004, as quoted in Fader-Hardie note 031 (academic)\strut
\end{minipage} & \begin{minipage}[t]{0.30\columnwidth}\raggedright
count framing\strut
\end{minipage} & \begin{minipage}[t]{0.30\columnwidth}\raggedright
models ``for estimating the number of active customers have been proposed \ldots{} (Schmittlein, Morrison, and Columbo 1987)''\strut
\end{minipage}\tabularnewline
\bottomrule
\end{longtable}

Counter-evidence, in fairness: the R BTYD and CLVTools documentation are careful that PAlive is a point-in-time quantity ``at the end of the calibration period'' (the horizon reading is added downstream); the lifetimes issue tracker and the PyMC-Marketing quickstart both flag the BG/NBD's P(alive) = 1 artifact for one-time buyers; and two tutorials caution that their constant-P(alive) horizon assumptions are crude. None of these caveats concerns the category error itself, which is the gap this paper fills. Full URLs and access records accompany the replication materials.

\end{document}